\documentclass{aa}

\usepackage{graphicx}
\usepackage{txfonts}
\usepackage{lipsum}
\usepackage{subcaption}         
\usepackage{lscape}             
\usepackage{placeins}          
\usepackage{natbib}

\bibpunct{(}{)}{;}{a}{}{,}
\usepackage{amsmath}
\usepackage{xcolor} 
\usepackage[normalem]{ulem}
\usepackage[hyperindex,breaklinks]{hyperref}
\hypersetup{
    colorlinks=true,              
    linkcolor=blue,               
    citecolor=blue,               
    urlcolor=blue                 
}

\begin{document}

   \title{Robustness of pairwise kinematic Sunyaev--Zel'dovich effect to optical-cluster-selection bias}

   \subtitle{}

   \author{Y.-H. Hsu\inst{1, 2, 3}\thanks{yuhhsu@asiaa.sinica.edu.tw, orcid: 0000-0003-0381-562X}
        \and D. Gruen\inst{1, 4}
        \and P. A. Gallardo\inst{5}
        \and K. Dolag\inst{1, 6}
        \and C.-H. To\inst{7}
        \and H.-Y. Wu\inst{8}
        \and I. Marini\inst{9}
        \and E. Rozo\inst{10}
        }

   \institute{University Observatory, Faculty of Physics, Ludwig-Maximilians-Universit\"{a}t, Scheinerstr. 1, 81679 Munich, Germany\\
            \and Max-Planck-Institut f\"{u}r extraterrestrische Physik (MPE), Giessenbachstrasse 1, 85748 Garching bei M\"{u}nchen, Germany\\
            \and Institute of Astronomy and Astrophysics, Academia Sinica  (ASIAA),  Taipei 10617, Taiwan\\
            \and Excellence Cluster ORIGINS, Boltzmannstr. 2, 85748 Garching, Germany\\
            \and Department of Physics and Astronomy, University of Pennsylvania, Philadelphia, PA 19104, USA\\
            \and Max-Planck-Institut f\"{u}r Astrophysik, Karl-Schwarzschild-Strasse 1, 85741 Garching, Germany\\
            \and Department of Astronomy and Astrophysics, University of Chicago, Chicago, Illinois 60637, USA\\
            \and Department of Physics, Southern Methodist University, Dallas, Texas 75205, USA\\
            \and European Southern Observatory, Karl Schwarzschildstrasse 2, 85748, Garching bei M\"{u}nchen, Germany\\
            \and Department of Physics, University of Arizona, 1118 East Fourth Street, Tucson, AZ 85721, USA\\
            }

   \date{Accepted April 1, 2026}

   \abstract{The pairwise kinematic Sunyaev--Zel'dovich\,(kSZ) effect measures both the pairwise motion between galaxy groups and clusters and the amount of gas within them, providing a tracer for cosmic growth. To interpret the cosmological information in the kSZ measurements, it is crucial to understand the optical-cluster-selection bias on the kSZ observables. Line-of-sight structures that contribute to both the optical observable (e.g.,~richness) and the cosmological signal can induce a correlation between these two quantities at a fixed cluster mass. The selection bias arising from this correlation is a key systematic effect for cosmological analyses. For cosmological observables such as cluster abundance and weak lensing, controlling this selection bias may help explain the tension between the DES-Y1 results and the Planck constraints. In order to test for a kSZ effect equivalent of such a bias, we adopted an alternative mock richness based on galaxy counts within cylindrical volumes along the line of sight. We applied the cylindrical count method to hydrodynamical simulations across a wide range of galaxy-selection criteria, assigning richness consistent with DES-Y1 to the mock clusters. When comparing optically selected clusters to mass-selected halos, we find no significant bias on pairwise kSZ signals, pairwise velocities, or optical depth within our uncertainty limits of approximately 16, 10, and 8 \%, respectively.
   }


   \maketitle
\nolinenumbers

\section{Introduction} \label{sec:intro}
During the growing process of large-scale structure, matter flows toward gravitational potential wells, eventually forming dense objects such as galaxy clusters. Measurements of the motions of material therefore provide direct insight into cosmic growth and offer constraints on cosmological physics -- including dark energy -- that are complementary to those from matter-density tracers.

Although direct spectroscopic observation of the peculiar velocity of galaxy clusters is challenging due to cosmic expansion, several methods have been developed to extract this information. The pairwise kinematic Sunyaev--Zel'dovich\,(kSZ) effect is one of the cosmic-growth tracers\,\citep{Ferreira1999Streaming, Hand2012EvidenceEffect}.
The kSZ effect is a velocity-induced change in the cosmic microwave background\,(CMB) temperature, arising from the interaction of CMB photons with the bulk motion of ionized gas in the intra-cluster medium\,\citep[ICM;][]{Sunyaev1980TheMeasurement}. The signal is proportional to the line-of-sight peculiar velocity and the number density of the electrons, providing an approach for measuring the peculiar motion and the distribution of baryons. Two clusters tend to move toward each other due to gravity, leading to a net CMB temperature difference for the pairs, which is known as the pairwise kSZ signal. The pairwise kSZ signal can be expressed as the product of the mean optical depth, $\bar{\tau},$ of the cluster sample, and the mean relative pairwise velocity, $v_{12}$\,\citep{Hand2012EvidenceEffect, Soergel2016DetectionSPT}, 
\[T_{pkSZ}(r) \simeq 
\bar{\tau} v_{12}(r)\frac{T_{\mathrm{CMB}}}{c}\,,\]
where $T_{pkSZ}$ is the net deviation on the temperature of the CMB blackbody spectrum, $T_{\mathrm{CMB}}$.
The pairwise kSZ effect provides a tool for distinguishing between models of acceleration of the cosmic expansion rate and for constraining the cosmic total neutrino mass\,\citep{Mueller, Mueller2014ConstraintsEffect}. The dependence of pairwise kSZ signals on the cosmic growth rate, $f,$ and the present-day root-mean-square of matter density fluctuations, $\sigma$, is $f\sigma^{2}$. This is complementary to other growth tracers such as redshift space distortions,  which depend on $f\sigma$.

The kSZ signals also trace the distribution of baryons associated with dark-matter halos via the optical depth, $\tau$, which is highly sensitive to the baryonic feedback processes in galaxy groups and cluster halos. The amount of baryonic feedback is a crucial systematic in modern weak lensing cosmology, especially in evaluating the significance of a potential tension between $S_8$ measurements\,\citep{Chisari2019ModellingCosmology, McCarthy2024FLAMINGO:Structure, Bigwood2024WeakFeedback}. Recent analyses comparing stacked kSZ profiles of massive galaxies with hydrodynamical simulations suggest a scenario with more aggressive feedback than previously assumed\,\citep{Hadzhiyska2024EvidenceGalaxies, Bigwood2024WeakFeedback, McCarthy2024FLAMINGO:Structure}.

Since the initial measurement by \citet{Hand2012EvidenceEffect} of the pairwise kSZ effect, the detection significance has reached a level of more than 9$\sigma$\,\citep{Calafut2021TheGalaxies, Chen2022, Li2024DetectionSpace, Hadzhiyska2025, Gong2025}. Most past works cross-matched optical large surveys with CMB images, taking advantage of the large sample size of optical galaxies. There are two main approaches to identify proxies for the centers of massive groups and clusters: one assumes that a significant fraction of luminous galaxies are group and cluster centrals\,\citep{Hand2012EvidenceEffect, PlanckCollaboration2015, DeBernardis2017DetectionTelescope, Sugiyama2018, Calafut2021TheGalaxies, Hadzhiyska2025, Gong2025}, while the other applies group- and cluster-finding algorithms\,\citep{Soergel2016DetectionSPT, Chen2022, Schiappucci2023MeasurementDes, Li2024DetectionSpace}. Throughout this paper, we refer to these algorithm-selected systems as ``optical clusters,'' although their mass range includes galaxy groups and clusters. Compared to luminous galaxies, algorithm-selected clusters represent the high-mass end of the halo mass function. Massive clusters have the highest pairwise velocities and optical depths\,\citep{Mueller2014ConstraintsEffect}. Moreover, clusters dominate the impact on weak lensing of baryonic effects\,\citep{To2024DecipheringClusters}, highlighting the importance of kSZ measurements using massive cluster samples with well-defined masses. However, while the luminous galaxy samples can be fully spectroscopically selected, the cluster samples are photometrically selected and have limited spectroscopic coverage.

The pairwise kSZ measurement on optical clusters is entering a new stage with the latest spectroscopic surveys. Large spectroscopic surveys such as the Dark Energy Spectroscopic Instrument\,(DESI) significantly enhance the spectroscopic coverage of optical clusters, paving the way to significantly detect the pairwise kSZ signal on fully spectroscopically supported cluster center samples\,\citep{Levi2013The2013, DESICollaboration2025DataInstrument}. Linear theory predicts pairwise velocity profiles that peak at comoving separations of approximately 20--50\,Mpc and approach zero beyond 300\,Mpc. With photometric data, it is challenging to measure the peak of the signal on small scales\,($\leq$ 50\,Mpc) due to the suppression of the signal by photometric redshift uncertainties\,\citep{Soergel2016DetectionSPT}. Measurement attempts on spectroscopically supported optical cluster samples have not reached significant detection due to the limited sample size as compared to other tracers\,\citep{DeBernardis2017DetectionTelescope, Chen2022}. Large spectroscopic surveys therefore have the potential to greatly improve the constraining power of the kSZ signal. This strengthens the demand for better control of systematic effects on kSZ observables. However, the optical-selection effect on kSZ signals has not yet been quantified.

Cluster-selection bias arises from the correlation between the cluster mass proxy used for selection and the other observables of cosmological interest, such as weak lensing or surrounding galaxy density. If not properly modeled, this bias can lead to systematic errors in halo mass estimation and affect the inferred cosmology. For instance, selection bias likely contributes to the tension between the cosmological constraints from Planck and from the DES-Y1 optical cluster abundances and the small-scale gravitational lensing signal\,\citep{Abbott2018DarkLensing, Abbott2020DarkLensing, To2021CombinationSurvey, Costanzi2021CosmologicalData}.
This is because the total mass of galaxy clusters is not directly observable, and cluster identification relies on mass proxies. The mass proxy richness, $\lambda$, a weighted sum of the counts of member galaxies, is widely used to select clusters in optical surveys\,\citep{Rykoff2014RedMaPPer.CATALOG, Rykoff2016TheData}. However, since the selection is based on photometry, objects along the line of sight can be misidentified as member galaxies\,\citep{Costanzi2018ModelingCatalogues, Myles2021SpectroscopicCatalogue}. Additionally, halo orientation and concentration could contribute to this bias, as cluster finders typically assume halos are spherically symmetric\,\citep{Osato2018, Zhang2022IncorporatingAnalyses}. This can lead to the tendency of halos with overestimated richness also having higher values of other observables with respect to the mean value at their halo mass. In other words, after removing the mass dependence, the residual of richness remains correlated with the residuals of the other observables. Quantifying this effect is important for interpreting the cosmological information from the cluster properties. Although modern spectroscopic surveys provide redshifts for central galaxies and a substantial number of bright members, the cluster samples are still defined photometrically and therefore remain subject to the same selection effects. Moreover, the limited spectroscopic completeness prevents the construction of a purely spectroscopic cluster sample with a comparable size and redshift coverage, implying that understanding selection bias will remain critical for fully exploiting the statistical power of optical cluster catalogs.

Optical-selection bias is thus a critical systematic uncertainty in modern cluster cosmology. Several models have been developed to address the impact of optical selection bias on the weak lensing signal\,\citep{Sunayama2020, Wu2022OpticalLensing}, with the potential to resolve the tension among the cosmological constraints from DES-Y1 cluster counts, lensing, and \textit{Planck}\,\citep{Salcedo2024ConsistencyiPlanck/i}.

In this work, we extended the method of \citet{Wu2022OpticalLensing}, which studied the optical-selection effect on mock galaxy and cluster catalogs generated from N-body simulations, to hydrodynamical simulations. We applied this approach to investigate the optical-cluster-selection bias on pairwise kSZ signals and its components -- pairwise velocity and optical depth. Hydrodynamical simulations are essential for this purpose, because studying the selection bias on kSZ signals requires detailed modeling of the gas distribution in the line-of-sight large-scale structures. Although peculiar velocity is not a projected quantity, pairwise velocity measurements could still be affected by line-of-sight structures, for example when clusters are preferentially aligned with a filament. This will pave the way for confident interpretation of upcoming pairwise kSZ measurements and shed light on the gas distribution in galaxy clusters. 

This paper is structured as follows. In Sect.\,\ref{sec:formula}, we briefly present the formulas of the pairwise kSZ effect. Sections\,\ref{sec:sim} and\,\ref{sec:methods} detail the simulations and the method. Section\,\ref{sec:discuss} presents the results with a brief discussion. Section\,\ref{sec:sum} summarizes the paper. Following the simulations, we assumed the WMAP7 $\Lambda$CDM cosmology\,\citep{Komatsu2011SEVEN-YEARINTERPRETATION}, with a Hubble constant of $H_0=\mathrm{70.4\,km\, s}^{-1}\mathrm{Mpc}^{-1}$; matter density of $\Omega_\mathrm{m}=0.272$; cosmological constant, $\Omega_\Lambda=0.728;$  normalization of the power spectrum, $\sigma_8 = 0.809$; and spectral index of the primordial power spectrum, $n_s = 0.963$. The halo masses are presented in $\mathrm{log\,M_{200m}} (h^{-1}\mathrm{M_{\odot}}$) if not specified. Physical and comoving distances are denoted as pMpc and cMpc, respectively.

\section{Formalism}\label{sec:formula}

\subsection{Pairwise kSZ effect}\label{subsec:pkSZ}
The kSZ effect can be observed by a small deviation, $\Delta T_{\mathrm{kSZ}}$, in the temperature of the CMB blackbody spectrum, $T_{\mathrm{CMB}}$\,\citep{Sunyaev1980TheMeasurement}:

\[\frac{\Delta T_{\mathrm{kSZ}}}{T_{\mathrm{CMB}}} = -\sigma_T \int{dl\,n_e(r) \frac{\mathbf{\hat{r} \cdot v_e}(r)}{c}}\,,\]
where $\sigma_T$ is the Thomson cross section; $l$ is the line-of-sight distance; $n_e$ and $\mathbf{v_e}$ are the number density and peculiar velocity of the electrons in a cluster in the direction of the unit vector, $\mathbf{\hat{r}}$; and $c$ is the speed of light. According to the definition of optical depth,

\[ \tau=\sigma_T\int dl\,n_e(r) \,,\]
and if we ignore the internal motion, the kSZ effect is simplified to

\[\frac{\Delta T_{\mathrm{kSZ}}}{T_{\mathrm{CMB}}} \simeq -\tau\frac{v_{\textrm{\footnotesize los}}}{c}\,.\]
If we assume that there is no strong correlation between $\tau$ and $\lvert v_{\mathrm{los}}\rvert$,

\[-\langle T_1-T_2\rangle \simeq 
\frac{T_{\mathrm{CMB}}}{c}\langle \tau_1 v^{\textrm{\footnotesize los}}_1 - \tau_2 v^{\textrm{\footnotesize los}}_2 \rangle
\simeq
\bar{\tau} v_{12}(r)\frac{T_{\mathrm{CMB}}}{c}\,,\]
where $\bar{\tau}$ is the average optical depth of the cluster sample. The assumptions implied by this equation are tested on a cluster sample in hydro-simulation\,\citep{Soergel2018CosmologySimulations}.

\subsection{Pairwise velocity}\label{subsec:v12}
The pairwise velocity profile $v_{12}(r,z)$,  depending on the 3D pair separation and redshift, is sensitive to the growth of cosmic structure. Linear theory shows

\[ v_{12}(r,z) \simeq -\frac{2}{3}arHf_g\frac{b\bar{\xi}(r,a)}{1+b^{2}\xi(r,a)}\,,\] 
where $f_g$ is the cosmic growth rate, $H$ is the Hubble parameter at the scale factor $a$, and $\xi$ is the two-point correlation function. The halo correlation function $\xi_h$ is related to $\xi$ through the mass-averaged halo bias $b$, as the approximation $\xi_h \simeq b^2 \xi$\,. $\bar{\xi}$ is the averaged correlation function within a comoving radius, r\,(see \citealt{Soergel2016DetectionSPT} and \citealt{Mueller} for detailed derivations).

\subsection{Optical-cluster-selection bias} \label{subsec:selbias}
The optical-cluster-selection effect occurs when there is a conditional dependence between the richness residual $\Delta\mathrm{ln}\lambda$ and the residual of another observable $\Delta\mathrm{ln} D_\mathrm{obs}$ at a given mass:

\[P(\Delta\mathrm{ln}\lambda, \Delta \mathrm{ln}D_\mathrm{obs}|\mathrm{ln}M) \neq P(\Delta\mathrm{ln}\lambda|\mathrm{ln}M)P(\Delta \mathrm{ln}D_\mathrm{obs}|\mathrm{ln}M)\] 
\[\Delta\mathrm{ln}\lambda = \mathrm{ln}\lambda-\langle \mathrm{ln}\lambda|\mathrm{ln}M \rangle\]
\[\Delta \mathrm{ln}D_\mathrm{obs} = \mathrm{ln}D_\mathrm{obs}-\langle \mathrm{ln}D_\mathrm{obs}|\mathrm{ln}M \rangle,\]
where $\lambda$ is richness, defined as in the red-sequence matched-filter probabilistic percolation (\verb|redMaPPer|) cluster finder\,\citep{Rykoff2014RedMaPPer.CATALOG}. The cluster mass is denoted by $M$, and $D_\mathrm{obs}$ represents one of the cluster observables on which the bias might emerge, such as the kSZ effect or weak gravitational lensing. $\Delta\mathrm{ln}\lambda$ and $\Delta \mathrm{ln}D_\mathrm{obs}$ represent the residuals of $\lambda$ and the observable on the logarithmic space after removing their mass dependence, typically modeled by a power-law relation (see \citealt{Wu2022OpticalLensing} for a bivariate Gaussian model for $\mathrm{ln}\lambda$ and the lensing residual).

\section{Simulations}\label{sec:sim}
To investigate the selection effect on kSZ signals, it is essential to accurately model the distribution of baryonic gas within the large-scale structures where halos are embedded. In this work, we used the hydrodynamical simulations of \verb|Magneticum|\footnote{\href{http://www.magneticum.org/}{http://www.magneticum.org/}}\,\citep{Dolag2025EncyclopediaDay}. \verb|Magneticum| is a series of cosmological simulations that follow WMAP7 cosmology, including a comprehensive range of physical processes, such as a subgrid model of radiative cooling, a uniform time-dependent UV background, star formation and stellar feedback, growth of black holes, and feedback from active galactic nuclei\,(AGNs), which are essential for modeling the ICM\,\citep{Hirschmann2014CosmologicalDownsizing}. The subgrid models are calibrated to reproduce the gas properties of clusters at $z=0$. Remarkably, even without tuning to stellar properties, the simulations reproduce many stellar-based scaling relations\,\citep{Dolag2025EncyclopediaDay}, while it could be challenging for cosmological simulations calibrated primarily to stellar properties to accurately model gas scaling relations and kinematic properties\,\citep{Popesso2024, vandeSande2019}. Moreover, \verb|Magneticum| is able to reproduce the critical properties of SZ observations from Planck\,\citep{Dolag2016, Dolag2025EncyclopediaDay}. Therefore, the \texttt{Magneticum} simulations are particularly suitable for studying the ICM.

Several SZ products are constructed from \verb|Magneticum|, such as light cones and full sky maps\,\citep{Soergel2018CosmologySimulations, Dolag2016, Coulton2022EffectsStudies}. In this work, we analyzed a kSZ map along with the associated cluster and galaxy catalogs from a 5$\times$5 $\mathrm{deg}^2$ light cone with a depth of $z<2.1$, generated from Box2 of \texttt{Magneticum} by the \texttt{SMAC} code\,\citep{Dolag2005ThePlanck}\footnote{\href{https://wwwmpa.mpa-garching.mpg.de/~kdolag/Smac/}{https://wwwmpa.mpa-garching.mpg.de/~kdolag/Smac/}}. Box2 has a size of 352 $h^{-1}\mathrm{cMpc}$ and a mass of dark-matter particles of $6.9\times 10^8$ $h^{-1}\mathrm{M_{\odot}}$, which allowed us to resolve galaxies. The light cone is approximated by 27 redshift slices taken from random positions in the simulation box. This is because interpolating positions of gas particles between snapshots is challenging, especially for large hydrodynamical simulations, when the snapshots are relatively far from each other. The full depth kSZ map is the co-addition of the maps calculated from the redshift slices. The geometry of the z=$0.2-0.6$ light cone is described in Table \ref{table:lc}. The light cone provides multiwavelength mock data, including X-ray emission\,\citep{Biffi2018AGNSimulations}, SZ Compton Y, the kSZ effect, and optical magnitudes (see \citealt{Marini2024DetectingEra, Marini2025DetectingSurveys} for light-cone design). The stellar mass function in the simulation is complete down to $10^{9.8}\,\mathrm{M_{\odot}}$. The combination of comprehensive ICM physics and high resolution makes these simulations well suited for studying selection effects on kSZ measurements. We also verified that the mock sky coordinates from the flat light cone are sufficiently accurate for the pairwise kSZ analysis\,(see Appendix \ref{appendix:lc_coord}). In addition, we used clusters from a 35$\times$35 $\mathrm{deg}^2$ light cone\,\citep{Soergel2018CosmologySimulations} to generate an empirical pairwise velocity template\,(see Appendix\,\ref{appendix:MF}). This larger light cone is based on Box0, which has moderate resolution and greater volume, making it suitable for capturing large-scale velocity statistics and resolving massive galaxy clusters.

\begin{table}
\begin{center}
\caption{ Geometry and the redshift slices of the z=$0.2-0.6$ region in the 5$\times$5\, $\mathrm{deg^2}$ light cone.}
\begin{tabular}{ccc}
\hline
  \multicolumn{1}{c}{Redshift} &
  \multicolumn{1}{c}{Depth\,[cMpc]} &
  \multicolumn{1}{c}{Width\,[cMpc]} \\
\hline 
   0.19\,$<$\,z\,$<$\,0.23 & 151 & 76\\
  0.23\,$<$\,z\,$<$\,0.27 & 152 & 89\\
  0.28\,$<$\,z\,$<$\,0.32 & 154 & 103\\
  0.32\,$<$\,z\,$<$\,0.36 & 156 & 116\\
  0.36\,$<$\,z\,$<$\,0.40 & 157 & 130\\
  0.40\,$<$\,z\,$<$\,0.45 & 158 & 144\\
  0.45\,$<$\,z\,$<$\,0.50 & 159 & 158\\
  0.50\,$<$\,z\,$<$\,0.54 & 160 & 172\\
  0.54\,$<$\,z\,$<$\,0.59 & 161 & 186\\
  0.59\,$<$\,z\,$<$\,0.65 & 161 & 200\\
\hline\end{tabular}
\label{table:lc}
\end{center}
\end{table}

\section{Methods} \label{sec:methods}

This section outlines the key steps in the investigation into optical-selection bias on kSZ observables across a wide range of galaxy-selection criteria. Section\,\ref{subsec:richness} details how galaxies were selected based on physical properties and how the cylindrical count method was applied to each set of the selection criteria to mimic the optical-cluster-selection process that assigns richness to each cluster. Section\,\ref{subsec:kSZ and components} describes how the observables -- pairwise kSZ effect, pairwise velocity, and optical depth -- were measured on the mock data products. 
Finally, Sect.\,\ref{subsec:reconstruct} presents the estimation of selection bias by comparing signals of richness-selected halos with signals of equal mass distribution sets of halos selected purely by their mass.

\subsection{Alternative richness} \label{subsec:richness}

\begin{figure*}
\sidecaption
\includegraphics[width=12cm]{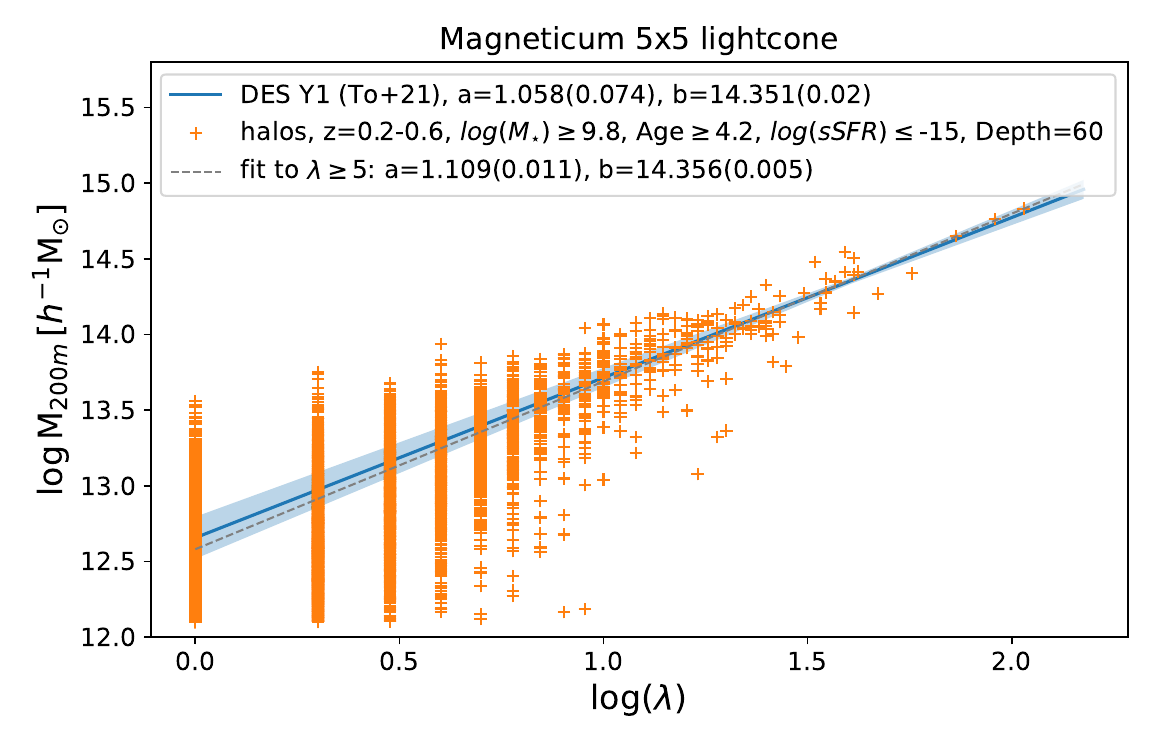}
 \caption{$M\,-\,\lambda$ relation of halos in the Magneticum 5$\times$5 $\mathrm{deg^2}$ light cone, given a galaxy selection of $\mathrm{M_{\star}}\geq 10^{9.8}\,\mathrm{M_{\odot}}$, mean stellar age\,$\geq 4.2\,$Gyr, $\mathrm{sSFR}\leq 10^{-15}\,\mathrm{yr^{-1}}$, and cylindrical depth of 60 \,$h^{-1}\mathrm{cMpc}$. The relation fit to $\lambda\geq5$ (dashed line) is consistent with DES-Y1 results within $1\sigma$\,\citep[][blue line and shaded region]{To2021DarkCorrelations}.}
 \label{fig:lambda-M}
\end{figure*}

To mimic the optical-cluster-selection process, we implemented the method of cylindrical counts from \citet{Wu2022OpticalLensing}. The method defines a cylindrical region at the position of each cluster, with the depth of the cylinder corresponding to the photometric redshift uncertainty along the line of sight, and the aperture corresponding to the cluster size. The counts of galaxies in the cylinders serve as the alternative richness. The alternative richness accounts for the projection effect and avoids the color criteria of richness, which are challenging for simulations. The cylindrical counts have been shown to reproduce the selection effect that biases the weak lensing signal\,\citep{Sunayama2020, Wu2022OpticalLensing, Salcedo2024ConsistencyiPlanck/i}.

Here we describe the procedures for assigning alternative richness to each cluster. First, we selected subhalos in the light cone based on their stellar mass, average stellar age, and specific star formation rate (sSFR) to obtain a ``bright'' and ``red'' galaxy sample.  These properties use all particles that are bound to each subhalo identified by the \texttt{SUBFIND} algorithm\,\citep{Springel2001Populating0, Dolag2009SubstructuresSimulations}. We define bright and red by the physical properties of all particles in each subhalo, because the goal is not to generate galaxy properties that are directly comparable with observations, but to explore a wide range of galaxy properties consistent with the $M$ - $\lambda$ relation of optical clusters. The stellar mass cut includes the redshift dependence of the characteristic magnitude, $m^*(z),$ of the Schechter luminosity function as \verb|redMaPPer|\,\citep{Rykoff2016TheData}. We adopted the $m^*(z)$ derived for the SDSS $r$-band from the model of \citet{Bruzual2003Stellar2003}
\footnote{\href{https://github.com/erykoff/redmapper/blob/main/how-to/README.md}{https://github.com/erykoff/redmapper/blob/main/how-to/README.md}}, with a $(1+z)$ evolution from the $M/L$ ratio of elliptical galaxies, as considered in \citet{Saglia2010TheEvolution}. The minimum value of each stellar mass cut, $M_{\star}(z),$ is set at $z=0.65$, and $M_{\star}(z)$ increases as z decreases. The stellar mass function in the simulation is complete down to $10^{9.8}\,\mathrm{M_{\odot}}$.

Second, we selected the clusters at $z=0.2-0.6$ to compare with DES results, with a minimum mass of $\mathrm{log}\,(M_{200m}/h^{-1}\mathrm{M_{\odot}}) \geq 12.1$ to ensure the stellar masses of central galaxies were well resolved. For an accurate measurement of kSZ aperture photometry, clusters within 5\,arcmin of the edge of the light-cone map were excluded. These criteria yielded a sample of 23\,128 mock clusters. For simplicity, these mock catalogs assumed that the clusters to have the same locations and velocities as their central galaxies. In reality, among clusters, optically selected central galaxies, and gas, there could be offsets of positions and velocities, which is another effect to model in cosmological inferences\,\citep{Calafut2017ClusterExtraction, Zhang2019DarkCatalogues, Orlowski-Scherer2021AtacamaACT, Ding2024MiscenteringCounterparts}. This setup mimics the case of cross-matching optically selected central galaxies with spectroscopic redshift catalogs. Under this framework, the uncertainty in spectroscopic redshifts is negligible compared to other sources of uncertainty.

Third, we applied the cylindrical count pipeline to these two catalogs to assign richness to the clusters. We used $R_{200m}$ as an approximation for the size of the cluster defined by \verb|redMaPPer,| $R_{\lambda}$. The radius $R_{200m}$ tends to be larger than $R_{\lambda}$ for high-mass clusters and falls below it for low-mass clusters. \citet{Wu2022OpticalLensing} showed that using an approximate aperture might increase the bias, as $R_{\lambda}$ minimizes the scatter of X-ray luminosity at a fixed $\lambda$. 
If a galaxy is within the cylinders of multiple halos, it is considered to be a member of the most massive halo. The cylinder depth is defined as the maximum distance that can separate galaxies from the cluster center, and we used a range of 40--100\,cMpc to account for photometric redshift uncertainty in recent surveys\,\citep[the typical value is 50\,cMpc for DES; see][]{Soergel2016DetectionSPT}. Subsequently, we compared the best-fit $M_{200m}$ - $\lambda$ relation at $\lambda \geq 5$ to the DES-Y1 relation\,\citep{To2021DarkCorrelations}: 

\[\langle M_{200m}|\lambda \rangle=10^b 
\left(\frac{\lambda}{40}\right)^a
\,h^{-1}\mathrm{M_{\odot}}.\]
This comparison examines whether a set of galaxy-selection criteria is ruled out by observational data\,(Fig.\,\ref{fig:lambda-M}).

\subsection{Estimating the pairwise kSZ signal and its components}\label{subsec:kSZ and components}

In this subsection, we investigate the pairwise kSZ observable and its components: pairwise velocity and optical depth. The measurement pipeline of pairwise kSZ effect and pairwise velocity is the updated version of \verb|iskay|\footnote{\href{https://github.com/patogallardo/iskay}{https://github.com/patogallardo/iskay}}\,\citep{Gallardo2019OptimizingTelescope, Gallardo2025}.
This tool has been utilized to analyze the pairwise kSZ effect on ACT data\,\citep{Calafut2021TheGalaxies}. It can measure aperture photometry on the kSZ maps and measure the pairwise kSZ effect by the estimator of \citet{Ferreira1999Streaming},

\[ P(r) = - \frac{\Sigma{(\delta T_{i}-\delta T_j)c_{ij}}}{\Sigma{c_{ij}^2}}
\;,\quad 
c_{ij}=\hat{r}_{ij}\cdot\frac{{\hat{r}_i+\hat{r}}_j}{2} \,,\]
where $\delta T_i$ is the kSZ temperature decrement for cluster $i$. Each cluster pair is separated by a comoving distance of $r=|\mathbf{r}_{ij}|=|\mathbf{r}_i-\mathbf{r}_j|$, and $c_{ij}$ is the geometry weighting regarding how the pair is aligned with the line of sight. In other words, $c_{ij}=\mathrm{cos}\,\alpha$, where $\alpha$ is the angle between the pairs and the line of sight. The uncertainties for pairwise kSZ profiles were estimated by bootstrapping the halo catalog 1000 times.

The aperture photometry measures the average temperature, $T_{AP,}$ within a radius of $R_\theta$ and subtracts the average temperature from an outer annulus at $\sqrt{2}R_\theta$. We used apertures of radius $R_\theta=$2.1 and 2.7 arcmin, motivated by the size of the light-cone halos with $\mathrm{log}\,M_{200m} \geq 13.40\,h^{-1}\mathrm{M_{\odot}}$ (corresponding to $\lambda\geq5$ in DES-Y1). The median $R_{200m}$ for these halos at z=$0.2-0.6$ is 2.52 arcmin. For Legacy Survey \texttt{eROMaPPer} clusters with $\lambda\geq3.95$\,(which corresponds to $\lambda\geq5$ of DES-Y1, see \citet{Kluge2024TheSurvey}) at z=$0.01-1.0$, the median $R_{\lambda}$ is 2.19 arcmin. These sizes are also comparable to values used in previous studies based on luminous red galaxies\,\citep{Calafut2021TheGalaxies}.

In typical pairwise kSZ measurements, a correction is applied to aperture photometry to remove potential redshift-dependent contamination from the foreground or large-scale motion\,\citep{Hand2012EvidenceEffect}. This correction subtracts a Gaussian-smoothed average $\bar{T}_{AP}$ from fine redshift bins:
\[\bar{T}_{AP}(r_i, z_i, \sigma_z, R_\theta) = \frac{\Sigma_j T_{AP}(r_i, R_\theta)G(z_i, z_j, \sigma_z)}{G(z_i, z_j, \sigma_z)} \]
\[\delta T_i(r_i, z_i, \sigma_z, R_\theta) = T_{AP}(r_i, z_i, R_\theta)-\bar{T}_{AP}(r_i, z_i, \sigma_z, R_\theta)\].

The Gaussian function $G(z_i, z_j, \sigma_z) = \mathrm{exp}[-(z_i, z_j)^2/(2\sigma_z^2)]$ is controlled by the redshift smoothing parameter $\sigma_z$. In our main analysis, we did not apply this redshift-evolution correction. This decision is based on the absence of significant redshift evolution in the kSZ temperatures measured over $z = 0.2$–$0.6$ in the light cone. The mock images do not include foreground contamination, so there is no evolution directly from them, nor indirectly from the correlation between them and the measurement process. In addition, due to the limited size of the light cone, we found that it is challenging to properly measure a global redshift evolution systematic without significantly removing the local motion signal. Applying this correction to the light cone could in itself introduce bias. Hence, we disabled this function when using \texttt{iskay}.

The optical depth of the cluster gas, $\tau,$ is a key property that can be inferred from kSZ measurements. When fitting pairwise velocity templates with a fixed cosmology to pairwise kSZ data, the fit amplitude corresponds to the average $\tau$ of the sample. To estimate the potential selection bias that may propagate from kSZ signals to $\tau$, we estimated optical depth by fitting a simple linear regression line to $\Delta T_{kSZ}-v$ in the fine mass bins. The regression line introduced a $v^2$ weighting for each cluster, enhancing the signal compared to when directly averaging the ratio of $\Delta T_{kSZ}$ to $v$. This method does not introduce bias, as the average peculiar velocity in each bin is expected to be zero and independent of mass. We verified that the mean velocities in the mass bins were consistent with zero and that the $v^2$ weighting did not affect the main results. The uncertainty in $\tau$ for each mass bin was estimated by bootstrapping the halo catalog 1000 times.

\subsection{Reconstruction of the unbiased signal} \label{subsec:reconstruct}
\begin{figure}
\centering
    \includegraphics[width=0.5\textwidth]{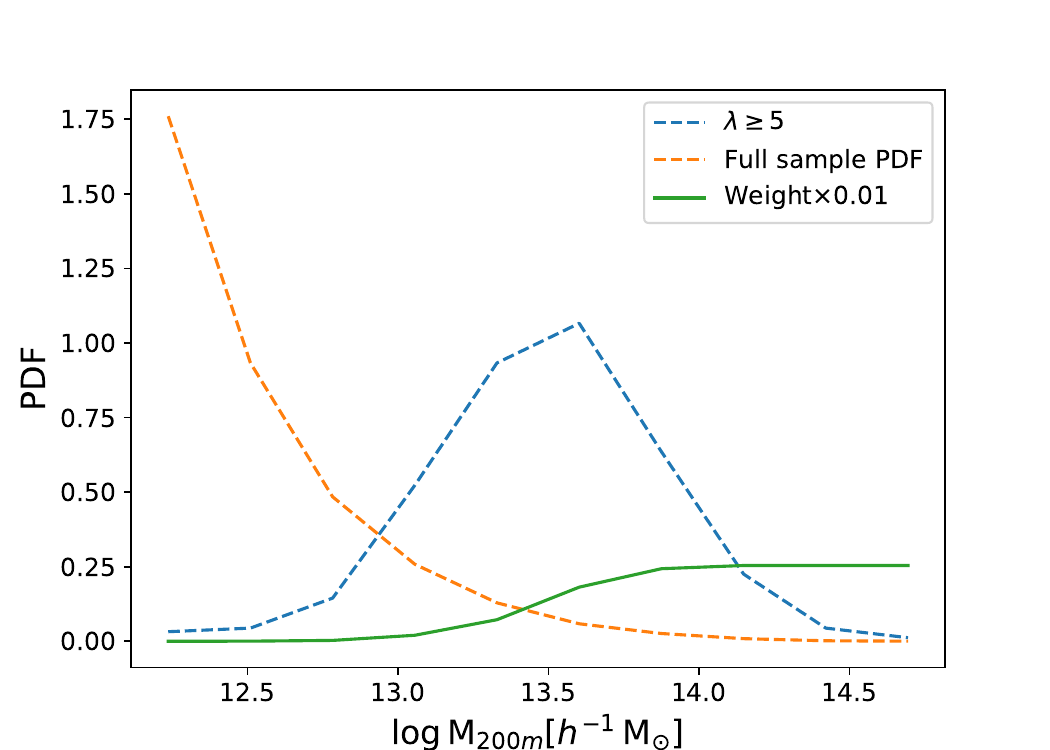}
    \caption{Mass distribution of the clusters with $\lambda \geq5$, shown as probability distribution function\,(dashed blue line). The galaxy-selection criteria used here are the same as in Fig.\,\ref{fig:lambda-M}. The dashed orange line represents the mass distribution of halos with $\mathrm{log}\,(M_{200m}/h^{-1}\mathrm{M_{\odot}}) \geq 12.1$ in the light cone. The green line indicates the weight applied to reconstruct the unbiased signal, scaled by a factor of 0.01 for visual comparison.}
    \label{fig:M_pdf}
\end{figure}

\begin{figure}
\centering
    \includegraphics[width=0.5\textwidth]{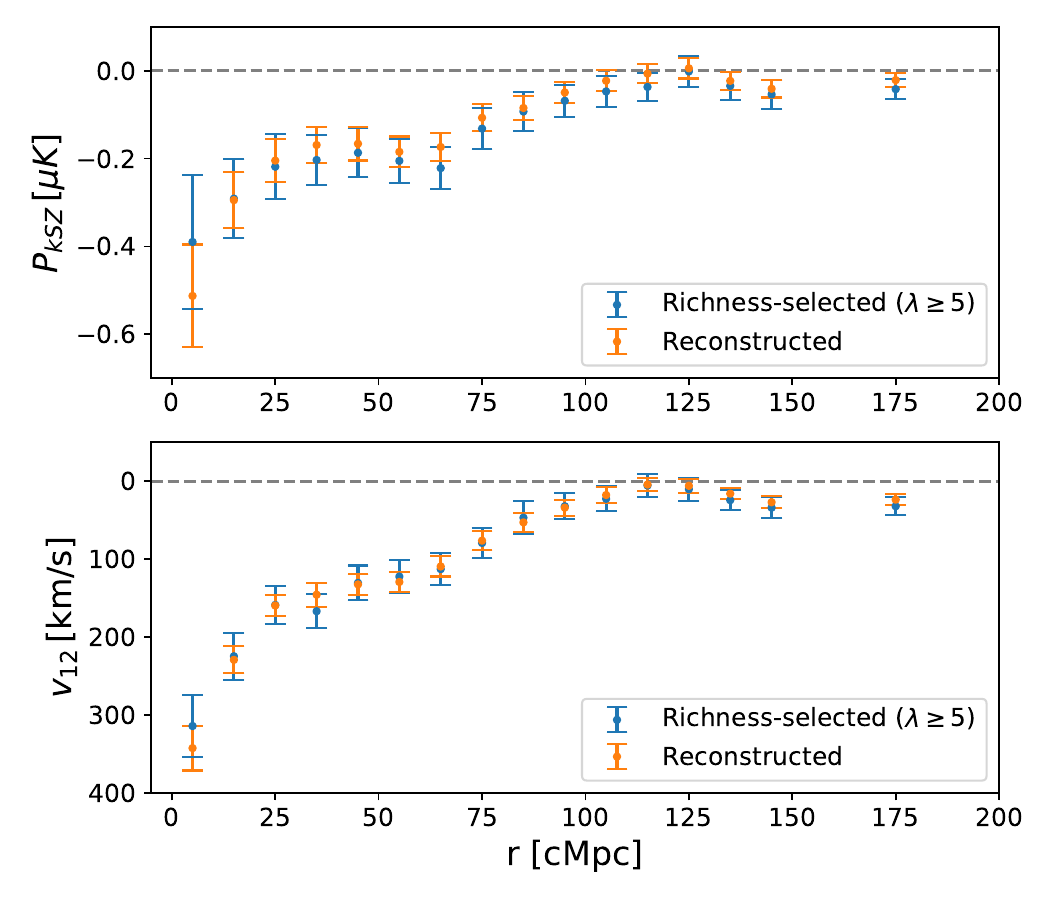}
    \caption{Comparison of pairwise signal of the ``richness-selected'' and the ``reconstructed'' sample of clusters. The richness-selected sample is potentially subject to selection bias, where the reconstructed sample is mass-selected, unbiased, and weighted to match the mass probability distribution of the richness-selected sample. Using the same galaxy selection as in Fig.\,\ref{fig:lambda-M}, the biases of pairwise kSZ effect, pairwise velocity, and optical depth are 1.10$\pm$0.19, 1.01$\pm$0.11, and 1.00$\pm$0.08.}
    \label{fig:eg_pairwsie}
\end{figure}
After obtaining the richness catalog, we studied the difference between the kSZ effect of a richness-selected sample and a reconstructed, unbiased signal. We used different approaches for pairwise observables and optical depth. The following describes the reconstruction of the pairwise kSZ effect and the pairwise velocities. A richness-selected sample was selected by richness cuts ($\lambda \geq 5$ in this work) and was subjected to selection bias. A reconstructed signal $P_{\mathrm{Recon}}$(r) was mass selected, but each cluster was weighted according to its mass to reconstruct the underlying mass distribution of the richness-selected sample\,(Fig.\,\ref{fig:M_pdf}):
\[ P_{\mathrm{Recon}}(r) = - \frac{\Sigma{w_i(M_i)w_j(M_j)(\delta T_{i}-\delta T_j)c_{ij}}}{\Sigma{w_i(M_i)w_j(M_j)c_{ij}^2}}. \]Namely, the weight, $w_i(M_i),$ represents the probability ratio that a cluster of a given mass will be selected by the richness cut to it being selected in the full halo catalog, given a specific galaxy selection. The bias value was estimated as the ratio between the richness-selected signal and the reconstructed unbiased signal\,(Fig.\,\ref{fig:eg_pairwsie} illustrates an example):
\[\mathrm{Bias=\frac{\mathbf{T}_{MF}\mathbf{P}_{kSZ}(\mathrm{\lambda\geq5})}{\mathbf{T}_{MF}\mathbf{P}_{kSZ,\,Recon}}}\].

We employed an optimized filter, $\mathbf{T}_\mathrm{MF}$, while averaging the pairwise profile over the separation of pairs, to optimize the S/N. The filter was constructed from the empirical signal and error template from the \texttt{Magneticum} $35\times35\,\mathrm{deg^2}$ light cone\,(see Appendix \ref{appendix:MF}).
The uncertainty of the richness-selected or reconstructed signal was estimated by summing their covariance matrix with the filter:
\[\mathrm{\sigma_{MF} = (\mathbf{T}_{MF}}^T\mathrm{\mathbf{C}_{BS}\mathbf{T}_{MF}})^{\frac{1}{2}}\,,\]
This uncertainty was then propagated to the final bias ratio.

On the other hand, regarding the reconstruction of optical depth, since this parameter is estimated on individual clusters, there is an opportunity to further reduce the uncertainties. We followed the ``weighting'' method in \citet{Wu2022OpticalLensing}, splitting the sample into ten mass bins, calculating the average $\tau$ in each bin\,(see Sect.\,\ref{subsec:kSZ and components}) and averaging the binned $\tau$ with the weighting of the mass distribution of the richness-selected sample\,(as Fig.\,\ref{fig:M_pdf}).

\section{Results and discussion} \label{sec:discuss}

\subsection{Selection bias}
\begin{figure*}
\sidecaption
  \includegraphics[width=12cm]{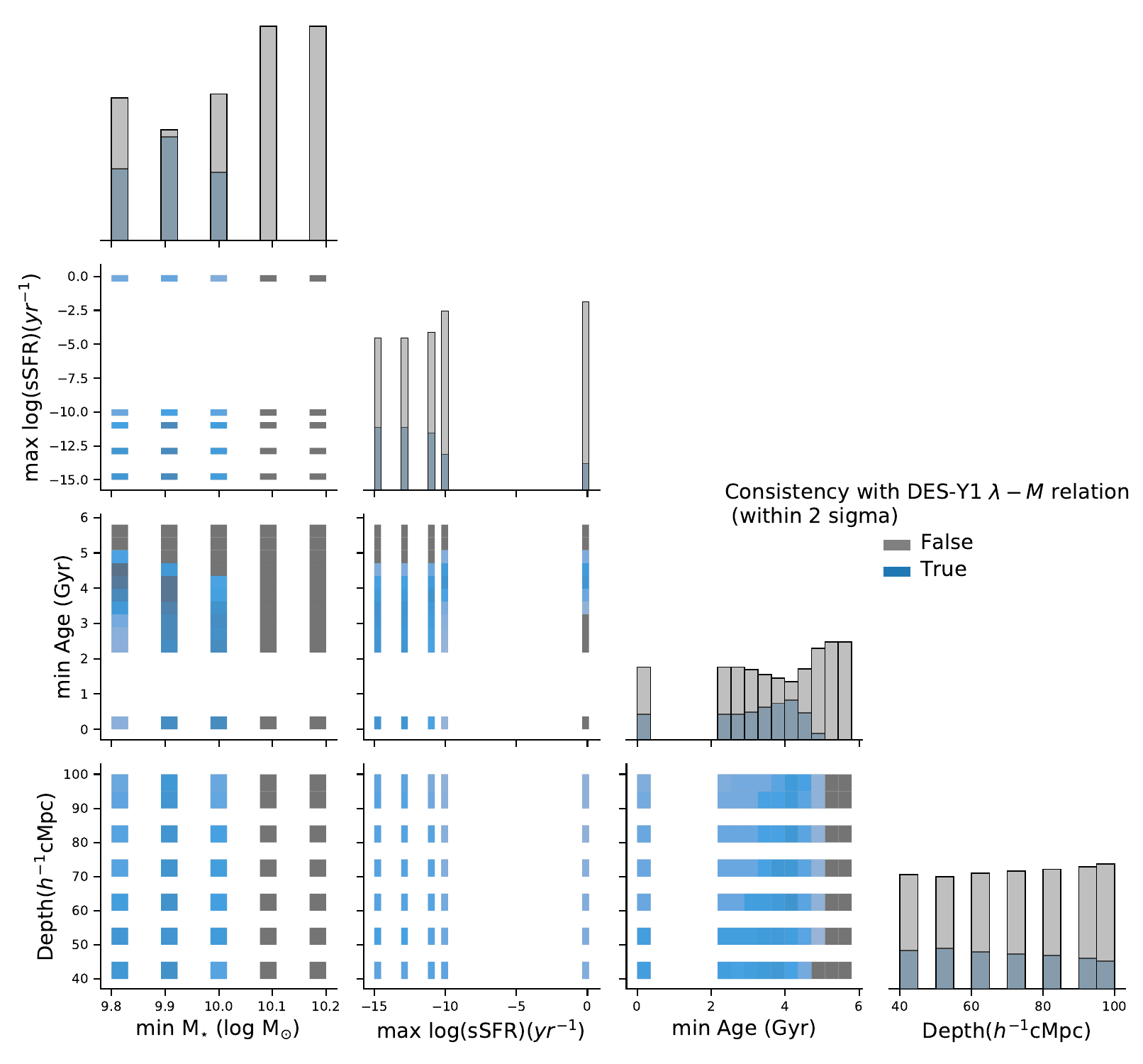}
     \caption{Galaxy-selection criteria explored to examine if the $M-\lambda$ relations are consistent with the DES-Y1 $M-\lambda$ relation within 2 $\sigma$. The blue squares are color-coded by the distribution of the selection criteria that generate $M-\lambda$ relations consistent with DES-Y1 across the full redshift range z$=0.2-0.6$ and within each of the four redshift bins. The ``min stellar mass'' denotes the minimum value set at $z=0.65$ for the redshift evolution of the stellar mass cut.}
     \label{fig:richness_2sigma}
\end{figure*}

Here we elaborate on the main results. In Fig.\,\ref{fig:eg_pairwsie}, we present an example with a specific galaxy selection, stellar mass $\geq 10^{9.8} \,\mathrm{M_{\odot}}$, mean stellar age\,$\geq 4.2\,$Gyr, $\mathrm{sSFR}\leq 10^{-15}\,\mathrm{yr^{-1}}$, and the cylindrical depth$=60 \,h^{-1}\mathrm{cMpc}$. The richness-selected signal and the reconstructed signal (defined in Sect.\,\ref{subsec:reconstruct}) show no significant differences, and their ratio is consistent with unity. For both the pairwise kSZ effect and the pairwise velocities, no bias is observed at the current level of uncertainties. In the full analysis, this measurement was repeated on a broad range of galaxy-selection criteria that generate $M-\lambda$ relations compatible with DES-Y1. The explorations of these selection criteria are shown in Fig.\,\ref{fig:richness_2sigma}. Here the ``min stellar mass'' refers to the threshold applied at $z=0.65$\,(see Sect.\,\ref{subsec:richness} for the redshift evolution of the stellar mass cut). The grid displays the range of galaxy selections and cylindrical depths explored. Each combination produces a $M-\lambda$ relation. The grids are color-coded according to the distribution of the $M-\lambda$ relations that are compatible with DES-Y1. We examined both extremes of the galaxy bimodality by controlling the sSFR and mean stellar age. We used the most restrictive ``red'' criterion, which includes passive galaxies down to modest stellar mass, and the ``bluest'' criterion, which effectively removes the color criterion and only selects galaxies by stellar mass. The cylindrical depth and the stellar mass cut balance the ``color'' criteria, generating reasonable $M-\lambda$ relations. The $M-\lambda$ relations were compared with DES-Y1, on the full redshift range z$=0.2-0.6$, and within four redshift bins of width 0.1. 

\begin{figure*}
\centering
    \includegraphics[width=\textwidth]{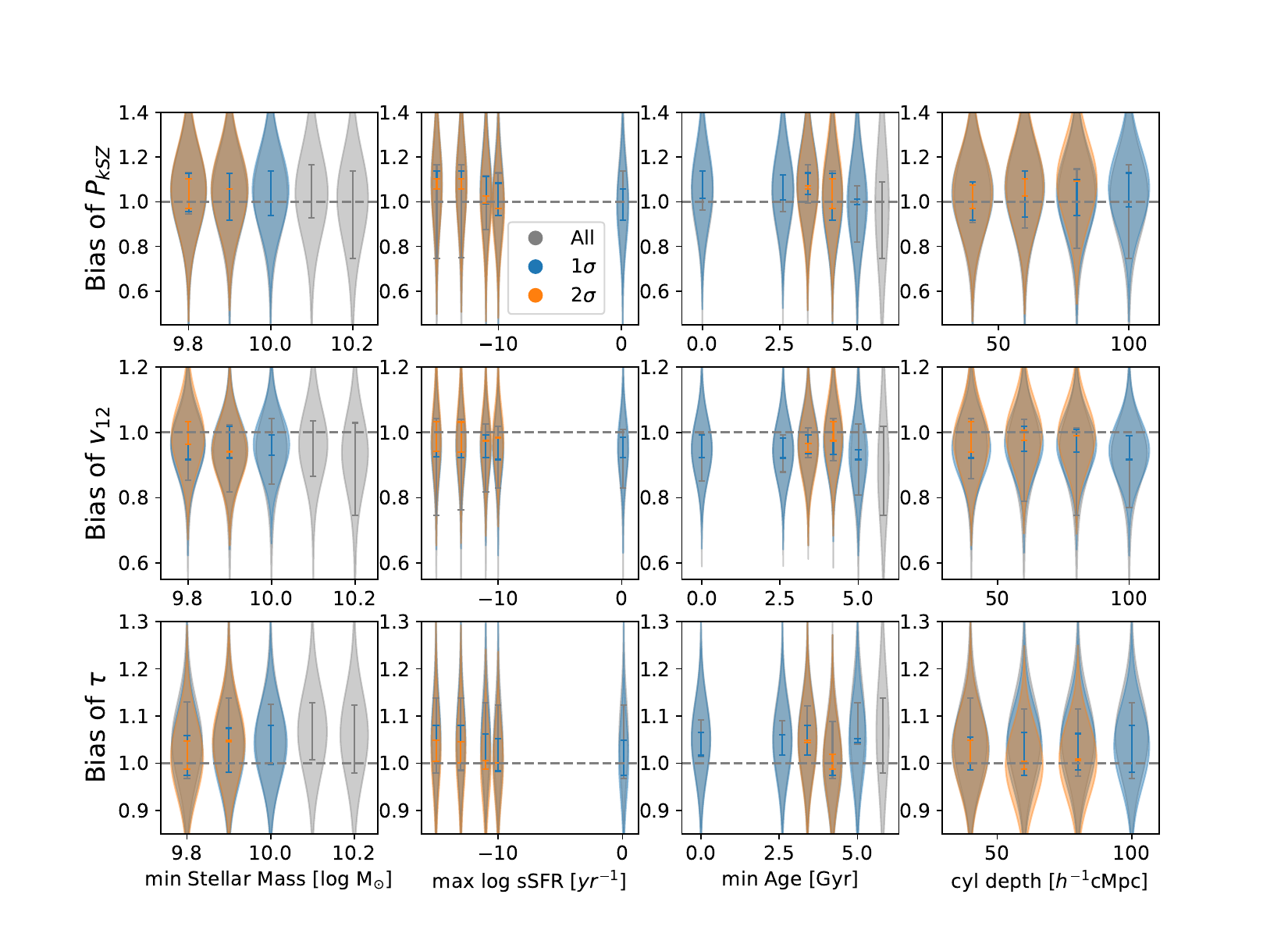}
    \caption{Constraints on bias values based on a smoothing scale of $R_{\theta}=2.7$ arcmin and all galaxy-selection criteria explored\,(gray). The violin areas include both statistical errors and systematic errors within a set of proposed galaxy selections.  Each bootstrap uncertainty is considered as a Gaussian kernel extended to 3$\sigma$. The widths of the violins are normalized to avoid overlapping. Bars indicate the range of bias values for each fixed parameter. The horizontal dashed lines mark the case of no bias\,(Bias$=$1). The biases derived from $M-\lambda$ relations consistent with DES-Y1\,\citep{To2021DarkCorrelations} within 1$\sigma$\,(2$\sigma$) are shown in orange\,(blue). For $R_{\theta}=2.7$ arcmin, and with galaxy selections consistent with DES-Y1 both across the full redshift range and within each of the four redshift bins of width 0.1, the biases are consistent with unity within 1$\sigma$. No significant bias is observed.}
    \label{fig:bias_is_one}
\end{figure*}

\begin{figure*}
\centering
    \includegraphics[width=\textwidth]{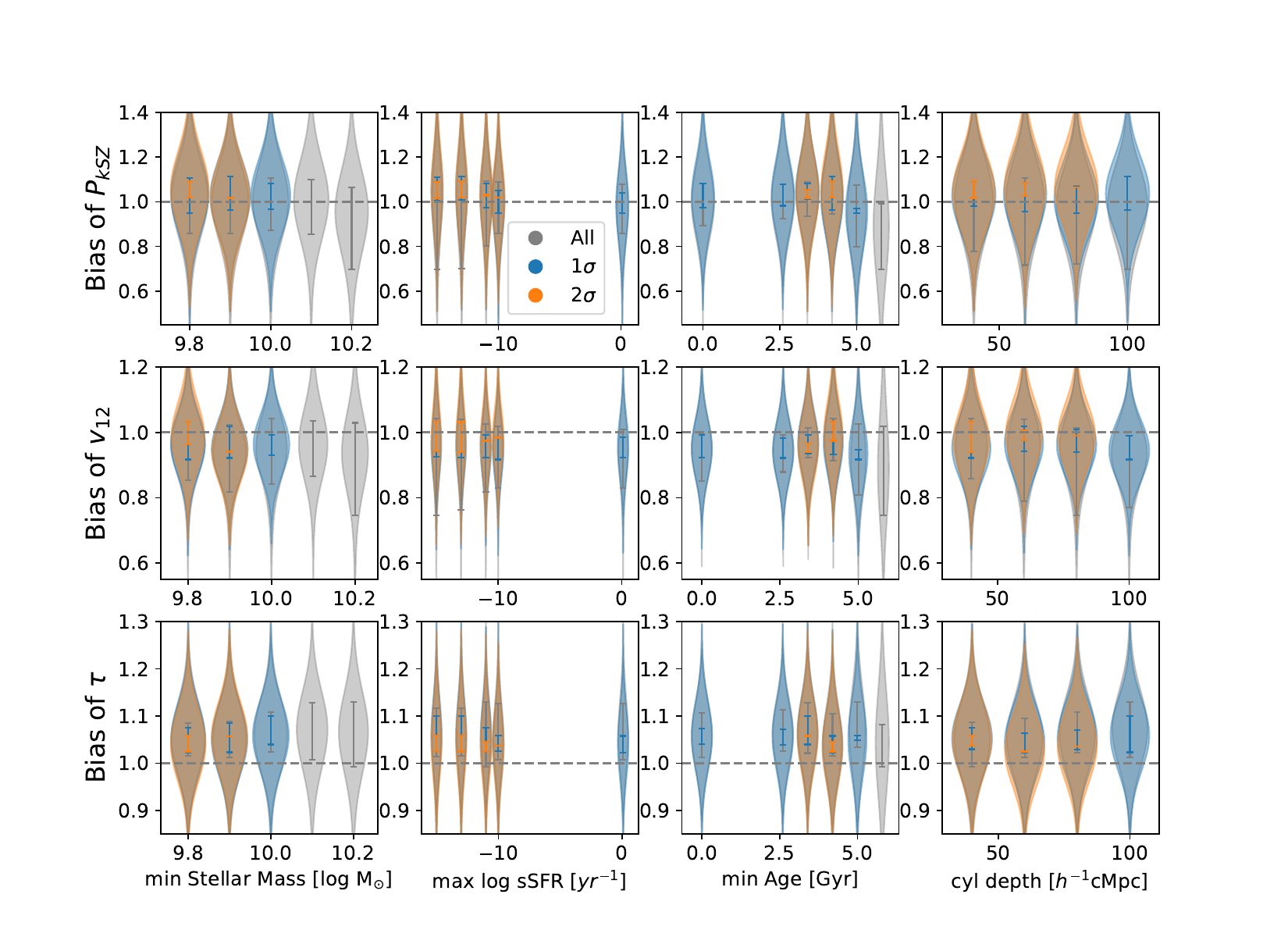}
    \caption{Similar to Fig.\,\ref{fig:bias_is_one} but for a smoothing scale of $R_{\theta}=2.1$ arcmin. For $R_{\theta}=2.1$ arcmin, and with galaxy selections consistent with DES-Y1 both across the full redshift range and within each of the four redshift bins of width 0.1, no significant bias is observed. The biases of the kSZ signal and velocities are consistent with unity within 1$\sigma$, and the biases of optical depth are within 1.34$\sigma$.}
    \label{fig:bias_is_one_AP2p1}
\end{figure*}

Based on the alternative richness, for each galaxy selection we investigated the biases of pairwise kSZ signals, pairwise velocity, and optical depth\,(Fig.\,\ref{fig:bias_is_one}). For selections that yield $M-\lambda$ relations consistent with DES-Y1 in the z$=0.2-0.6$ range and within each of the four redshift bins, the biases are shown in the blue (DES 2$\sigma$) and orange (DES 1$\sigma$) violin plots. All these points are consistent with unity within 1$\sigma$. Figure\,\ref{fig:bias_is_one} is based on a photometric aperture, $R_{\theta}=2.7'$, and Fig.\,\ref{fig:bias_is_one_AP2p1} shows a similar result for $R_{\theta}=2.1'$, where the biases remain consistent with unity within 1.34$\sigma$. For $R_{\theta}=2.7'$, the result indicates that there is no significant bias on pairwise kSZ signals, pairwise velocity, or optical depth above the levels of $19\%$, $11\%$, and  $9\%$, respectively. The median uncertainty levels across galaxy selections are $16\%$, $10\%$, and $8\%$, respectively.

\subsection{Discussion}

Figure\,\ref{fig:bias_is_one} shows that the biases are consistent with one, indicating that no significant bias is observed in pairwise kSZ signals or their related properties. This contrasts with weak lensing, where the optical-selection bias is known to introduce considerable bias\,\citep{Wu2022OpticalLensing, Salcedo2024ConsistencyiPlanck/i}. Unlike weak lensing, where the signal scales linearly with mass, the projected kSZ signal does not follow a simple linear relation with optical depth or mass along the line of sight. The contribution from peculiar velocities of structures along the line of sight, combined with pairwise orientation and averaging, leads to a more complicated projection effect. We note that these findings apply specifically to the pairwise kSZ estimator and pairwise velocity. They may not directly generalize to other kSZ extraction approaches such as those based on velocity reconstruction. The uncertainty level of our test ($\sim$19\%) is comparable to the measurement uncertainties in previous pairwise kSZ detections using optically selected clusters, which have reached a significance level of $\sim5\sigma$. At this level of statistical precision, it is reasonable to expect that optical-selection bias does not play a dominant role. However, achieving tighter constraints will require larger simulation volumes with comparable resolution to improve statistical power, which will be essential for interpreting the higher significance pairwise kSZ measurements anticipated from upcoming datasets.

We assumed that $\tau$ and $v$ have a negligible correlation when deriving the pairwise kSZ formula\,(Sect.\,\ref{subsec:pkSZ}). This assumption is verified by \citet{Soergel2018CosmologySimulations} on a cluster sample in \texttt{Magneticum}. However, we probed this relation down to a much lower mass. Given this assumption, the biases are expected to follow a similar relation, where the bias of $\tau$ will be equal to the ratio of bias of $P_{kSZ}$ and bias of $v_{12}$. Our main result in Fig.\,\ref{fig:bias_is_one} aligns with this assumption. 
Further investigation of this relationship and the $\tau-v$ correlation would require a larger simulation sample.

\begin{figure*}
\centering
    \includegraphics[width=\textwidth]{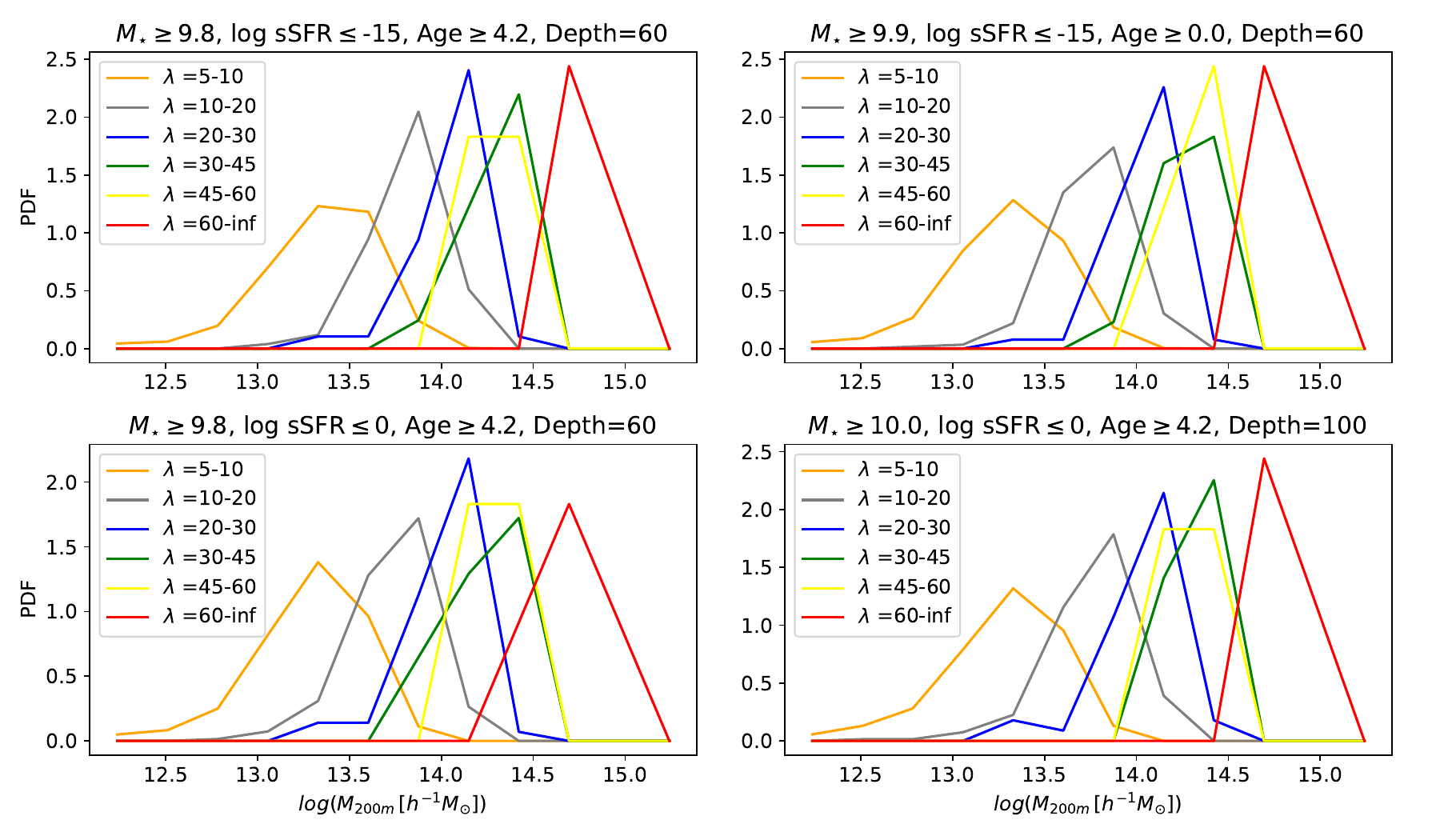}
    \caption{Mass distributions in richness bins [5, 10, 20, 30, 45, 60, inf], given different galaxy-selection criteria that have $M-\lambda$ consistent with DES-Y1. The upper left panel is based on a relatively restrictive passive selection, and the other panels are examples of less restrictive criteria. The less restrictive criteria make the distributions wider and generate stronger low-mass tails.}
    \label{fig:m_scatter}
\end{figure*}

We also investigated the mass probability distribution functions in the richness bins [5, 10, 20, 30, 45, 60, inf]. We find that less restrictive stellar age or sSFR criteria result in wider distributions and heavier low-mass tails\,(Fig.\,\ref{fig:m_scatter}). Compared with the distributions from \citet{Salcedo2024ConsistencyiPlanck/i}, which are based on a halo-occupation-distribution framework, the distributions in this work have fewer objects in the low-mass tail. Although the low-mass tail of the $\lambda > 20$ bins suffered from low number counts, fewer low-mass halos for a given richness implies less potential of selection bias. Future quantitative research for selection bias on the kSZ effect would require a larger simulation and a more sophisticated bias model to account for the $M-\lambda$ relation in fine bins.

Due to the limited size of the light cone, large-scale profiles (roughly beyond 200 cMpc) are highly uncertain. Furthermore, the \verb|Magneticum| light cones are constructed by slices centered at random positions of the box, and neither the mass nor velocity is correlated across the slices. The slice sizes are comparable with the depth of the cylinder\,(see Table\,\ref{table:lc}), which could reduce large-scale structure correlations with the clusters. However, these limitations affect both the richness-selected and the reconstructed samples in the same way. It would still be valuable to apply the same analysis on a larger light cone with a different construction.

The correction of redshift evolution\,(\ref{subsec:kSZ and components}) is not applied in this work, as no significant redshift evolution is detected in the mock kSZ temperatures, and the limited size of the light cone makes the proper correction challenging. In future studies that include foregrounds, this correction will need to be enabled. We do not expect the redshift evolution correction to introduce additional selection bias, because it does not directly depend on richness. However, foregrounds such as residuals from the thermal SZ effect may correlate with richness and are worth testing. In principle, thermal SZ residuals primarily affect the most massive clusters, which are rare and less sensitive to the selection bias, so they are not expected to generate a significant additional bias.

\section{Summary}\label{sec:sum}

The detection of the pairwise kSZ effect has now reached a significance level of more than 9$\sigma$. With the ongoing and upcoming large spectroscopic surveys, the constraining power of the kSZ effect on a scale of less than 50 cMpc is going to improve substantially. Robust modeling of systematics is essential to unlock this cosmological information.

In this work, we investigated the optical-cluster-selection bias on the pairwise kSZ effect, pairwise velocities, and optical depth. We applied the method of cylindrical counts to obtain reasonable $M-\lambda$ relations in hydrodynamical simulations. We estimated the bias for $\lambda \geq 5$ on clusters at z=$0.2-0.6$ with $\mathrm{M_{200m}} \geq 10^{12.1}\,h^{-1}\mathrm{M_{\odot}}$ over a wide range of galaxy-selection criteria. Within our uncertainty level, no significant bias was observed.

This result paves the way for upcoming pairwise kSZ measurements, with optical cluster samples powered by DESI spectroscopy, cross-matching with Atacama Cosmology Telescope\,\citep[ACT,][]{Coulton2024, Naess2025}, Simons Observatory\,\citep{Abitbol2025}, and next-generation CMB efforts. 
Applying this framework to larger hydrodynamical simulations, ideally incorporating various feedback or cosmological models, and a light-cone structure complementary to \verb|Magneticum|, would be highly valuable. For future work, a more sophisticated model would be required for a quantitative estimation, and investigating the bias of multi-observable probes---such as clustering and weak lensing---would provide deeper insights into the gas properties in large structures.

\begin{acknowledgements}
    The authors would like to thank Luca Sala, Stephan Vladutescu-Zopp, Frederick Groth, Johannes Stoiber, and Rhea-Silvia Remus for valuable discussions and assistance with accessing \texttt{Magneticum} simulations and related tools. We also thank Andr\'{e}s Salcedo and Tomomi Sunayama for insightful discussions on optical selection effects, as well as Justin Myles and Oliver Friedrich for helpful discussions. This work was funded by the Deutsche Forschungsgemeinschaft (DFG, German Research Foundation) under Germany's Excellence Strategy – EXC-2094/2 – 390783311. IM acknowledges support from the European Research Council (ERC) under the European Union’s Horizon Europe research and innovation programme ERC CoG (Grant agreement No. 101045437). KD acknowledges support by the COMPLEX project from the European Research Council (ERC) under the European Union’s Horizon 2020 research and innovation program grant agreement ERC-2019-AdG 882679. The calculations for the Magneticum simulations were carried out at the Leibniz Supercomputer Center (LRZ) under the project pr83li.
\end{acknowledgements}

\bibliographystyle{bibtex/aa} 
\bibliography{main}

@article{Abbott2018DarkLensing,
    title = {{Dark Energy Survey year 1 results: Cosmological constraints from galaxy clustering and weak lensing}},
    year = {2018},
    journal = {Physical Review D},
    author = {Abbott, T. M. C. and Abdalla, F. B. and Alarcon, A. and Aleksi{\'{c}}, J. and Allam, S. and Allen, S. and Amara, A. and Annis, J. and Asorey, J. and Avila, S. and Bacon, D. and Balbinot, E. and Banerji, M. and Banik, N. and Barkhouse, W. and Baumer, M. and Baxter, E. and Bechtol, K. and Becker, M. R. and Benoit-L{\'{e}}vy, A. and Benson, B. A. and Bernstein, G. M. and Bertin, E. and Blazek, J. and Bridle, S. L. and Brooks, D. and Brout, D. and Buckley-Geer, E. and Burke, D. L. and Busha, M. T. and Campos, A. and Capozzi, D. and Carnero Rosell, A. and Carrasco Kind, M. and Carretero, J. and Castander, F. J. and Cawthon, R. and Chang, C. and Chen, N. and Childress, M. and Choi, A. and Conselice, C. and Crittenden, R. and Crocce, M. and Cunha, C. E. and D’Andrea, C. B. and da Costa, L. N. and Das, R. and Davis, T. M. and Davis, C. and De Vicente, J. and DePoy, D. L. and DeRose, J. and Desai, S. and Diehl, H. T. and Dietrich, J. P. and Dodelson, S. and Doel, P. and Drlica-Wagner, A. and Eifler, T. F. and Elliott, A. E. and Elsner, F. and Elvin-Poole, J. and Estrada, J. and Evrard, A. E. and Fang, Y. and Fernandez, E. and Fert{\'{e}}, A. and Finley, D. A. and Flaugher, B. and Fosalba, P. and Friedrich, O. and Frieman, J. and Garc{\'{i}}a-Bellido, J. and Garcia-Fernandez, M. and Gatti, M. and Gaztanaga, E. and Gerdes, D. W. and Giannantonio, T. and Gill, M. S. S. and Glazebrook, K. and Goldstein, D. A. and Gruen, D. and Gruendl, R. A. and Gschwend, J. and Gutierrez, G. and Hamilton, S. and Hartley, W. G. and Hinton, S. R. and Honscheid, K. and Hoyle, B. and Huterer, D. and Jain, B. and James, D. J. and Jarvis, M. and Jeltema, T. and Johnson, M. D. and Johnson, M. W. G. and Kacprzak, T. and Kent, S. and Kim, A. G. and King, A. and Kirk, D. and Kokron, N. and Kovacs, A. and Krause, E. and Krawiec, C. and Kremin, A. and Kuehn, K. and Kuhlmann, S. and Kuropatkin, N. and Lacasa, F. and Lahav, O. and Li, T. S. and Liddle, A. R. and Lidman, C. and Lima, M. and Lin, H. and MacCrann, N. and Maia, M. A. G. and Makler, M. and Manera, M. and March, M. and Marshall, J. L. and Martini, P. and McMahon, R. G. and Melchior, P. and Menanteau, F. and Miquel, R. and Miranda, V. and Mudd, D. and Muir, J. and M{\"{o}}ller, A. and Neilsen, E. and Nichol, R. C. and Nord, B. and Nugent, P. and Ogando, R. L. C. and Palmese, A. and Peacock, J. and Peiris, H. V. and Peoples, J. and Percival, W. J. and Petravick, D. and Plazas, A. A. and Porredon, A. and Prat, J. and Pujol, A. and Rau, M. M. and Refregier, A. and Ricker, P. M. and Roe, N. and Rollins, R. P. and Romer, A. K. and Roodman, A. and Rosenfeld, R. and Ross, A. J. and Rozo, E. and Rykoff, E. S. and Sako, M. and Salvador, A. I. and Samuroff, S. and S{\'{a}}nchez, C. and Sanchez, E. and Santiago, B. and Scarpine, V. and Schindler, R. and Scolnic, D. and Secco, L. F. and Serrano, S. and Sevilla-Noarbe, I. and Sheldon, E. and Smith, R. C. and Smith, M. and Smith, J. and Soares-Santos, M. and Sobreira, F. and Suchyta, E. and Tarle, G. and Thomas, D. and Troxel, M. A. and Tucker, D. L. and Tucker, B. E. and Uddin, S. A. and Varga, T. N. and Vielzeuf, P. and Vikram, V. and Vivas, A. K. and Walker, A. R. and Wang, M. and Wechsler, R. H. and Weller, J. and Wester, W. and Wolf, R. C. and Yanny, B. and Yuan, F. and Zenteno, A. and Zhang, B. and Zhang, Y. and Zuntz, J.},
    number = {4},
    month = {8},
    pages = {043526},
    volume = {98},
    publisher = {American Physical Society},
    url = {https://link.aps.org/doi/10.1103/PhysRevD.98.043526},
    doi = {10.1103/PhysRevD.98.043526},
    issn = {2470-0010}
}

@article{Abbott2020DarkLensing,
    title = {{Dark Energy Survey Year 1 Results: Cosmological constraints from cluster abundances and weak lensing}},
    year = {2020},
    journal = {Physical Review D},
    author = {Abbott, T. M. C. and Aguena, M. and Alarcon, A. and Allam, S. and Allen, S. and Annis, J. and Avila, S. and Bacon, D. and Bechtol, K. and Bermeo, A. and Bernstein, G. M. and Bertin, E. and Bhargava, S. and Bocquet, S. and Brooks, D. and Brout, D. and Buckley-Geer, E. and Burke, D. L. and Carnero Rosell, A. and Carrasco Kind, M. and Carretero, J. and Castander, F. J. and Cawthon, R. and Chang, C. and Chen, X. and Choi, A. and Costanzi, M. and Crocce, M. and da Costa, L. N. and Davis, T. M. and De Vicente, J. and DeRose, J. and Desai, S. and Diehl, H. T. and Dietrich, J. P. and Dodelson, S. and Doel, P. and Drlica-Wagner, A. and Eckert, K. and Eifler, T. F. and Elvin-Poole, J. and Estrada, J. and Everett, S. and Evrard, A. E. and Farahi, A. and Ferrero, I. and Flaugher, B. and Fosalba, P. and Frieman, J. and Garc{\'{i}}a-Bellido, J. and Gatti, M. and Gaztanaga, E. and Gerdes, D. W. and Giannantonio, T. and Giles, P. and Grandis, S. and Gruen, D. and Gruendl, R. A. and Gschwend, J. and Gutierrez, G. and Hartley, W. G. and Hinton, S. R. and Hollowood, D. L. and Honscheid, K. and Hoyle, B. and Huterer, D. and James, D. J. and Jarvis, M. and Jeltema, T. and Johnson, M. W. G. and Johnson, M. D. and Kent, S. and Krause, E. and Kron, R. and Kuehn, K. and Kuropatkin, N. and Lahav, O. and Li, T. S. and Lidman, C. and Lima, M. and Lin, H. and MacCrann, N. and Maia, M. A. G. and Mantz, A. and Marshall, J. L. and Martini, P. and Mayers, J. and Melchior, P. and Mena-Fern{\'{a}}ndez, J. and Menanteau, F. and Miquel, R. and Mohr, J. J. and Nichol, R. C. and Nord, B. and Ogando, R. L. C. and Palmese, A. and Paz-Chinch{\'{o}}n, F. and Plazas, A. A. and Prat, J. and Rau, M. M. and Romer, A. K. and Roodman, A. and Rooney, P. and Rozo, E. and Rykoff, E. S. and Sako, M. and Samuroff, S. and S{\'{a}}nchez, C. and Sanchez, E. and Saro, A. and Scarpine, V. and Schubnell, M. and Scolnic, D. and Serrano, S. and Sevilla-Noarbe, I. and Sheldon, E. and Smith, J. Allyn. and Smith, M. and Suchyta, E. and Swanson, M. E. C. and Tarle, G. and Thomas, D. and To, C. and Troxel, M. A. and Tucker, D. L. and Varga, T. N. and von der Linden, A. and Walker, A. R. and Wechsler, R. H. and Weller, J. and Wilkinson, R. D. and Wu, H. and Yanny, B. and Zhang, Y. and Zhang, Z. and Zuntz, J.},
    number = {2},
    month = {7},
    pages = {023509},
    volume = {102},
    publisher = {American Physical Society},
    url = {https://link.aps.org/doi/10.1103/PhysRevD.102.023509},
    doi = {10.1103/PhysRevD.102.023509},
    issn = {2470-0010}
}

@article{Abitbol2025,
   author = {M. Abitbol and I. Abril-Cabezas and S. Adachi and P. Ade and A.E. Adler and P. Agrawal and J. Aguirre and Z. Ahmed and S. Aiola and T. Alford and A. Ali and D. Alonso and M.A. Alvarez and R. An and K. Arnold and P. Ashton and Z. Atkins and J. Austermann and S. Azzoni and C. Baccigalupi and A. Baleato Lizancos and D. Barron and P. Barry and J. Bartlett and N. Battaglia and R. Battye and E. Baxter and A. Bazarko and J.A. Beall and R. Bean and D. Beck and S. Beckman and J. Begin and A. Beheshti and B. Beringue and T. Bhandarkar and S. Bhimani and F. Bianchini and E. Biermann and S. Biquard and B. Bixler and S. Boada and D. Boettger and B. Bolliet and J.R. Bond and J. Borrill and J. Borrow and C. Braithwaite and T.L.R. Brien and M.L. Brown and S.M. Bruno and S. Bryan and R. Bustos and H. Cai and E. Calabrese and V. Calafut and F.M. Carl and A. Carones and J. Carron and A. Challinor and P. Chanial and N. Chen and K. Cheung and B. Chiang and Y. Chinone and J. Chluba and H.S. Cho and S.K. Choi and M. Chu and J. Clancy and S.E. Clark and P. Clarke and J. Cleary and D.L. Clements and J. Connors and C. Contaldi and G. Coppi and L. Corbett and N.F. Cothard and W. Coulton and K.D. Crowley and K.T. Crowley and A. Cukierman and J.M. D'Ewart and K. Dachlythra and R. Datta and S. Day-Weiss and T. de Haan and M. Devlin and L. Di Mascolo and S. Dicker and B. Dober and C. Doux and P. Dow and S. Doyle and C.J. Duell and S.M. Duff and A.J. Duivenvoorden and J. Dunkley and D. Dutcher and R. Dünner and M. Edenton and H. El Bouhargani and J. Errard and G. Fabbian and V. Fanfani and G.S. Farren and J. Fergusson and S. Ferraro and R. Flauger and A. Foster and K. Freese and J.C. Frisch and A. Frolov and G. Fuller and N. Galitzki and P.A. Gallardo and J.T. Galvez Ghersi and K. Ganga and J. Gao and X. Garrido and E. Gawiser and M. Gerbino and R. Gerras and S. Giardiello and A. Gill and V. Gilles and U. Giri and E. Gleave and V. Gluscevic and N. Goeckner-Wald and J.E. Golec and S. Gordon and M. Gralla and S. Gratton and D. Green and J.C. Groh and C. Groppi and Y. Guan and N. Gupta and J.E. Gudmundsson and S. Hagstotz and P. Hargrave and S. Haridas and K. Harrington and I. Harrison and M. Hasegawa and M. Hasselfield and V. Haynes and M. Hazumi and A. He and E. Healy and S.W. Henderson and B.S. Hensley and E. Hertig and C. Hervías-Caimapo and M. Higuchi and C.A. Hill and J.C. Hill and G. Hilton and M. Hilton and A.D. Hincks and G. Hinshaw and R. Hložek and A.Y.Q. Ho and S. Ho and S.P. Ho and T.D. Hoang and J. Hoh and E. Hornecker and A.L. Hornsby and S.C. Hotinli and Z. Huang and Z.B. Huber and J. Hubmayr and K. Huffenberger and J.P. Hughes and A. Idicherian Lonappan and M. Ikape and K. Irwin and J. Iuliano and A.H. Jaffe and B. Jain and H.T. Jense and O. Jeong and A. Johnson and B.R. Johnson and M. Johnson and M. Jones and B. Jost and D. Kaneko and E.D. Karpel and Y. Kasai and N. Katayama and B. Keating and B. Keller and R. Keskitalo and J. Kim and T. Kisner and K. Kiuchi and J. Klein and K. Knowles and A.M. Kofman and B.J. Koopman and A. Kosowsky and R. Kou and N. Krachmalnicoff and D. Kramer and A. Krishak and A. Krolewski and A. Kusaka and A. Kusiak and P. La Plante and A. La Posta and A. Laguë and J. Lashner and M. Lattanzi and A. Lee and E. Lee and J. Leech and C. Lessler and J.S. Leung and A. Lewis and Y. Li and Z. Li and M. Limon and L. Lin and M. Link and J. Liu and Y. Liu and J. Lonergan and T. Louis and T. Lucas and M. Ludlam and M. Lungu and M. Lyons and N. MacCrann and A. MacInnis and M. Madhavacheril and D. Mak and F. Maldonado and M. Mallaby-Kay and A. Manduca and A. Mangu and H. Mani and A.S. Maniyar and G.A. Marques and J. Mates and T. Matsumura and P. Mauskopf and A. May and N. McCallum and H. McCarrick and F. McCarthy and M. McCulloch and J. McMahon and P.D. Meerburg and Y. Mehta and J. Melin and J. Meyers and A. Middleton and A. Miller and M. Mirmelstein and K. Moodley and J. Moore and M. Morshed and T. Morton and E. Moser and T. Mroczkowski and M. Murata and M. Münchmeyer and S. Naess and H. Nakata and T. Namikawa and M. Nashimoto and F. Nati and P. Natoli and M. Negrello and S.K. Nerval and L. Newburgh and D.V. Nguyen and A. Nicola and M.D. Niemack and H. Nishino and Y. Nishinomiya and A. Orlando and J. Orlowski-Scherer and L. Pagano and L.A. Page and S. Pandey and A. Papageorgiou and I. Paraskevakos and B. Partridge and R. Patki and M. Peel and K. Perez Sarmiento and F. Perrotta and P. Phakathi and L. Piccirillo and E. Pierpaoli and T. Pinsonneault-Marotte and G. Pisano and D. Poletti and R. Puddu and G. Puglisi and F.J. Qu and M.J. Randall and C. Ranucci and C. Raum and R. Reeves and C.L. Reichardt and M. Remazeilles and Y. Rephaeli and D. Riechers and J. Robe and M.F. Robertson and N. Robertson and K. Rogers and F. Rojas and A. Romero and E. Rosenberg and A. Rotti and S. Rowe and A. Roy and S. Sadeh and N. Sailer and K. Sakaguri and T. Sakuma and Y. Sakurai and M. Salatino and G.H. Sanders and D. Sasaki and M. Sathyanarayana Rao and T.P. Satterthwaite and L. Saunders and L. Scalcinati and E. Schaan and B. Schmitt and M. Schmittfull and N. Sehgal and J. Seibert and Y. Seino and U. Seljak and S. Shaikh and E. Shaw and P. Shellard and B. Sherwin and M. Shimon and J.E. Shroyer and C. Sierra and J. Sievers and C. Sifón and P. Sikhosana and M. Silva-Feaver and S.M. Simon and A. Sinclair and K. Smith and W. Sohn and X. Song and R.F. Sonka and D. Spergel and J. Spisak and S.T. Staggs and G. Stein and J.R. Stevens and R. Stompor and E. Storer and R. Sudiwala and J. Sugiyama and K.M. Surrao and S. Sutariya and A. Suzuki and J. Suzuki and O. Tajima and S. Takakura and A. Takeuchi and I. Tansieri and A.C. Taylor and G. Teply and T. Terasaki and A. Thomas and D.B. Thomas and R. Thornton and H. Trac and T. Tsan and E. Tsang King Sang and C. Tucker and J. Ullom and L. Vacher and L. Vale and A. van Engelen and J. Van Lanen and J. van Marrewijk and D.D. Van Winkle and C. Vargas and E.M. Vavagiakis and I. Veenendaal and C. Vergès and M. Vissers and M. Viña and K. Wagoner and S. Walker and L. Walters and Y. Wang and B. Westbrook and J. Williams and P. Williams and H. Winch and E.J. Wollack and K. Wolz and J. Wong and Z. Xu and K. Yamada and E. Young and B. Yu and C. Yu and M. Zannoni and K. Zheng and N. Zhu and A. Zonca and I. Zubeldia},
   doi = {10.1088/1475-7516/2025/08/034},
   issn = {1475-7516},
   issue = {08},
   journal = {Journal of Cosmology and Astroparticle Physics},
   month = {8},
   pages = {034},
   title = {The Simons Observatory: science goals and forecasts for the enhanced Large Aperture Telescope},
   volume = {2025},
   url = {https://iopscience.iop.org/article/10.1088/1475-7516/2025/08/034},
   year = {2025}
}

@article{Biffi2018AGNSimulations,
    title = {{AGN contamination of galaxy-cluster thermal X-ray emission: Predictions for eRosita from cosmological simulations}},
    year = {2018},
    journal = {Monthly Notices of the Royal Astronomical Society},
    author = {Biffi, V. and Dolag, K. and Merloni, A.},
    number = {2},
    month = {12},
    pages = {2213--2227},
    volume = {481},
    publisher = {Oxford University Press},
    url = {https://ui.adsabs.harvard.edu/abs/2018MNRAS.481.2213B/abstract},
    doi = {10.1093/MNRAS/STY2436},
    issn = {13652966},
    arxivId = {1804.01096}
}

@article{Bigwood2024WeakFeedback,
    title = {{Weak lensing combined with the kinetic Sunyaev–Zel’dovich effect: a study of baryonic feedback}},
    year = {2024},
    journal = {Monthly Notices of the Royal Astronomical Society},
    author = {Bigwood, L and Amon, A and Schneider, A and Salcido, J and McCarthy, I G and Preston, C and Sanchez, D and Sijacki, D and Schaan, E and Ferraro, S and Battaglia, N and Chen, A and Dodelson, S and Roodman, A and Pieres, A and Fert{\'{e}}, A and Alarcon, A and Drlica-Wagner, A and Choi, A and Navarro-Alsina, A and Campos, A and Ross, A J and Rosell, A Carnero and Yin, B and Yanny, B and S{\'{a}}nchez, C and Chang, C and Davis, C and Doux, C and Gruen, D and Rykoff, E S and Huff, E M and Sheldon, E and Tarsitano, F and Andrade-Oliveira, F and Bernstein, G M and Giannini, G and Diehl, H T and Huang, H and Harrison, I and Sevilla-Noarbe, I and Tutusaus, I and Elvin-Poole, J and McCullough, J and Zuntz, J and Blazek, J and DeRose, J and Cordero, J and Prat, J and Myles, J and Eckert, K and Bechtol, K and Herner, K and Secco, L F and Gatti, M and Raveri, M and Kind, M Carrasco and Becker, M R and Troxel, M A and Jarvis, M and MacCrann, N and Friedrich, O and Alves, O and Leget, P -F and Chen, R and Rollins, R P and Wechsler, R H and Gruendl, R A and Cawthon, R and Allam, S and Bridle, S L and Pandey, S and Everett, S and Shin, T and Hartley, W G and Fang, X and Zhang, Y and Aguena, M and Annis, J and Bacon, D and Bertin, E and Bocquet, S and Brooks, D and Carretero, J and Castander, F J and da Costa, L N and Pereira, M E S and De Vicente, J and Desai, S and Doel, P and Ferrero, I and Flaugher, B and Frieman, J and Garc{\'{i}}a-Bellido, J and Gaztanaga, E and Gutierrez, G and Hinton, S R and Hollowood, D L and Honscheid, K and Huterer, D and James, D J and Kuehn, K and Lahav, O and Lee, S and Marshall, J L and Mena-Fern{\'{a}}ndez, J and Miquel, R and Muir, J and Paterno, M and Malag{\'{o}}n, A A Plazas and Porredon, A and Romer, A K and Samuroff, S and Sanchez, E and Cid, D Sanchez and Smith, M and Soares-Santos, M and Suchyta, E and Swanson, M E C and Tarle, G and To, C and Weaverdyck, N and Weller, J and Wiseman, P and Yamamoto, M},
    number = {1},
    month = {9},
    pages = {655--682},
    volume = {534},
    publisher = {Oxford University Press (OUP)},
    url = {https://academic.oup.com/mnras/article/534/1/655/7754165},
    doi = {10.1093/mnras/stae2100},
    issn = {0035-8711}
}

@article{Bruzual2003Stellar2003,
    title = {{Stellar population synthesis at the resolution of 2003}},
    year = {2003},
    journal = {MNRAS},
    author = {Bruzual, G. and Charlot, S.},
    number = {4},
    month = {10},
    pages = {1000--1028},
    volume = {344},
    url = {https://ui.adsabs.harvard.edu/abs/2003MNRAS.344.1000B/abstract},
    doi = {10.1046/J.1365-8711.2003.06897.X},
    issn = {0035-8711},
    arxivId = {arXiv:astro-ph/0309134},
    archivePrefix = {arXiv},
    eprint = {0309134},
    primaryClass = {astro-ph}
}

@article{Calafut2017ClusterExtraction,
    title = {{Cluster mislocation in kinematic Sunyaev-Zel’dovich effect extraction}},
    year = {2017},
    journal = {Physical Review D},
    author = {Calafut, Victoria and Bean, Rachel and Yu, Byeonghee},
    number = {12},
    month = {12},
    pages = {123529},
    volume = {96},
    publisher = {American Physical Society},
    url = {https://link.aps.org/doi/10.1103/PhysRevD.96.123529},
    doi = {10.1103/PhysRevD.96.123529},
    issn = {2470-0010}
}

@article{Calafut2021TheGalaxies,
    title = {{The Atacama Cosmology Telescope: Detection of the pairwise kinematic Sunyaev-Zel’dovich effect with SDSS DR15 galaxies}},
    year = {2021},
    journal = {Physical Review D},
    author = {Calafut, V. and Gallardo, P. A. and Vavagiakis, E. M. and Amodeo, S. and Aiola, S. and Austermann, J. E. and Battaglia, N. and Battistelli, E. S. and Beall, J. A. and Bean, R. and Bond, J. R. and Calabrese, E. and Choi, S. K. and Cothard, N. F. and Devlin, M. J. and Duell, C. J. and Duff, S. M. and Duivenvoorden, A. J. and Dunkley, J. and Dunner, R. and Ferraro, S. and Guan, Y. and Hill, J. C. and Hilton, G. C. and Hilton, M. and Hlo{\v{z}}ek, R. and Huber, Z. B. and Hubmayr, J. and Huffenberger, K. M. and Hughes, J. P. and Koopman, B. J. and Kosowsky, A. and Li, Y. and Lokken, M. and Madhavacheril, M. and McMahon, J. and Moodley, K. and Naess, S. and Nati, F. and Newburgh, L. B. and Niemack, M. D. and Page, L. A. and Partridge, B. and Schaan, E. and Schillaci, A. and Sif{\'{o}}n, C. and Spergel, D. N. and Staggs, S. T. and Ullom, J. N. and Vale, L. R. and Van Engelen, A. and Van Lanen, J. and Wollack, E. J. and Xu, Z.},
    number = {4},
    month = {8},
    pages = {043502},
    volume = {104},
    url = {https://link.aps.org/doi/10.1103/PhysRevD.104.043502},
    doi = {10.1103/PhysRevD.104.043502},
    issn = {2470-0010}
}

@article{Chen2022,
   author = {Ziyang Chen and Pengjie Zhang and Xiaohu Yang and Yi Zheng},
   doi = {10.1093/mnras/stab3604},
   issn = {0035-8711},
   issue = {4},
   journal = {Monthly Notices of the Royal Astronomical Society},
   month = {1},
   pages = {5916-5928},
   title = {Detection of pairwise kSZ effect with DESI galaxy clusters and Planck},
   volume = {510},
   url = {https://academic.oup.com/mnras/article/510/4/5916/6461109},
   year = {2022}
}

@article{Chisari2019ModellingCosmology,
    title = {{Modelling baryonic feedback for survey cosmology}},
    year = {2019},
    journal = {The Open Journal of Astrophysics},
    author = {Chisari, Nora Elisa and Mead, Alexander J. and Joudaki, Shahab and Ferreira, Pedro and Schneider, Aurel and Mohr, Joseph and Tr{\"{o}}ster, Tilman and Alonso, David and McCarthy, Ian G. and Martin-Alvarez, Sergio and Devriendt, Julien and Slyz, Adrianne and van Daalen, Marcel P.},
    number = {1},
    month = {5},
    volume = {2},
    publisher = {Maynooth University},
    url = {http://arxiv.org/abs/1905.06082 http://dx.doi.org/10.21105/astro.1905.06082},
    doi = {10.21105/astro.1905.06082},
    arxivId = {1905.06082v2}
}

@article{Costanzi2018ModelingCatalogues,
    title = {{Modeling projection effects in optically-selected cluster catalogues}},
    year = {2018},
    journal = {Monthly Notices of the Royal Astronomical Society},
    author = {Costanzi, M. and Rozo, E. and Rykoff, E. S. and Farahi, A. and Jeltema, T. and Evrard, A. E. and Mantz, A. and Gruen, D. and Mandelbaum, R. and DeRose, J. and McClintock, T. and Varga, T. N. and Zhang, Y. and Weller, J. and Wechsler, R. H. and Aguena, M.},
    number = {1},
    month = {7},
    pages = {490--505},
    volume = {482},
    publisher = {Oxford University Press},
    url = {http://arxiv.org/abs/1807.07072 http://dx.doi.org/10.1093/mnras/sty2665},
    doi = {10.1093/mnras/sty2665},
    arxivId = {1807.07072v1}
}

@article{Costanzi2021CosmologicalData,
    title = {{Cosmological constraints from DES Y1 cluster abundances and SPT multiwavelength data}},
    year = {2021},
    journal = {Physical Review D},
    author = {Costanzi, M. and Saro, A. and Bocquet, S. and Abbott, T. M. C. and Aguena, M. and Allam, S. and Amara, A. and Annis, J. and Avila, S. and Bacon, D. and Benson, B. A. and Bhargava, S. and Brooks, D. and Buckley-Geer, E. and Burke, D. L. and Carnero Rosell, A. and Carrasco Kind, M. and Carretero, J. and Choi, A. and da Costa, L. N. and Pereira, M. E. S. and De Vicente, J. and Desai, S. and Diehl, H. T. and Dietrich, J. P. and Doel, P. and Eifler, T. F. and Everett, S. and Ferrero, I. and Fert{\'{e}}, A. and Flaugher, B. and Fosalba, P. and Frieman, J. and Garc{\'{i}}a-Bellido, J. and Gaztanaga, E. and Gerdes, D. W. and Giannantonio, T. and Giles, P. and Grandis, S. and Gruen, D. and Gruendl, R. A. and Gupta, N. and Gutierrez, G. and Hartley, W. G. and Hinton, S. R. and Hollowood, D. L. and Honscheid, K. and James, D. J. and Jeltema, T. and Krause, E. and Kuehn, K. and Kuropatkin, N. and Lahav, O. and Lima, M. and MacCrann, N. and Maia, M. A. G. and Marshall, J. L. and Menanteau, F. and Miquel, R. and Mohr, J. J. and Morgan, R. and Myles, J. and Ogando, R. L. C. and Palmese, A. and Paz-Chinch{\'{o}}n, F. and Plazas, A. A. and Rapetti, D. and Reichardt, C. L. and Romer, A. K. and Roodman, A. and Ruppin, F. and Salvati, L. and Samuroff, S. and Sanchez, E. and Scarpine, V. and Serrano, S. and Sevilla-Noarbe, I. and Singh, P. and Smith, M. and Soares-Santos, M. and Stark, A. A. and Suchyta, E. and Swanson, M. E. C. and Tarle, G. and Thomas, D. and To, C. and Tucker, D. L. and Varga, T. N. and Wechsler, R. H. and Zhang, Z.},
    number = {4},
    month = {2},
    pages = {043522},
    volume = {103},
    publisher = {American Physical Society},
    url = {https://link.aps.org/doi/10.1103/PhysRevD.103.043522},
    doi = {10.1103/PhysRevD.103.043522},
    issn = {2470-0010}
}

@article{Coulton2022EffectsStudies,
    title = {{Effects of boosting on extragalactic components: methods and statistical studies}},
    year = {2022},
    journal = {Monthly Notices of the Royal Astronomical Society},
    author = {Coulton, William and Feldman, Sydney and Maamari, Karime and Pierpaoli, Elena and Yasini, Siavash and Dolag, Klaus},
    number = {2},
    month = {5},
    pages = {2252--2270},
    volume = {513},
    url = {https://academic.oup.com/mnras/article/513/2/2252/6570913},
    doi = {10.1093/mnras/stac1017},
    issn = {0035-8711}
}

@article{Coulton2024,
   author = {William Coulton and Mathew S. Madhavacheril and Adriaan J. Duivenvoorden and J. Colin Hill and Irene Abril-Cabezas and Peter A. R. Ade and Simone Aiola and Tommy Alford and Mandana Amiri and Stefania Amodeo and Rui An and Zachary Atkins and Jason E. Austermann and Nicholas Battaglia and Elia Stefano Battistelli and James A. Beall and Rachel Bean and Benjamin Beringue and Tanay Bhandarkar and Emily Biermann and Boris Bolliet and J. Richard Bond and Hongbo Cai and Erminia Calabrese and Victoria Calafut and Valentina Capalbo and Felipe Carrero and Grace E. Chesmore and Hsiao-mei Cho and Steve K. Choi and Susan E. Clark and Rodrigo Córdova Rosado and Nicholas F. Cothard and Kevin Coughlin and Kevin T. Crowley and Mark J. Devlin and Simon Dicker and Peter Doze and Cody J. Duell and Shannon M. Duff and Jo Dunkley and Rolando Dünner and Valentina Fanfani and Max Fankhanel and Gerrit Farren and Simone Ferraro and Rodrigo Freundt and Brittany Fuzia and Patricio A. Gallardo and Xavier Garrido and Jahmour Givans and Vera Gluscevic and Joseph E. Golec and Yilun Guan and Mark Halpern and Dongwon Han and Matthew Hasselfield and Erin Healy and Shawn Henderson and Brandon Hensley and Carlos Hervías-Caimapo and Gene C. Hilton and Matt Hilton and Adam D. Hincks and Renée Hložek and Shuay-Pwu Patty Ho and Zachary B. Huber and Johannes Hubmayr and Kevin M. Huffenberger and John P. Hughes and Kent Irwin and Giovanni Isopi and Hidde T. Jense and Ben Keller and Joshua Kim and Kenda Knowles and Brian J. Koopman and Arthur Kosowsky and Darby Kramer and Aleksandra Kusiak and Adrien La Posta and Victoria Lakey and Eunseong Lee and Zack Li and Yaqiong Li and Michele Limon and Martine Lokken and Thibaut Louis and Marius Lungu and Niall MacCrann and Amanda MacInnis and Diego Maldonado and Felipe Maldonado and Maya Mallaby-Kay and Gabriela A. Marques and Joshiwa van Marrewijk and Fiona McCarthy and Jeff McMahon and Yogesh Mehta and Felipe Menanteau and Kavilan Moodley and Thomas W. Morris and Tony Mroczkowski and Sigurd Naess and Toshiya Namikawa and Federico Nati and Laura Newburgh and Andrina Nicola and Michael D. Niemack and Michael R. Nolta and John Orlowski-Scherer and Lyman A. Page and Shivam Pandey and Bruce Partridge and Heather Prince and Roberto Puddu and Frank J. Qu and Federico Radiconi and Naomi Robertson and Felipe Rojas and Tai Sakuma and Maria Salatino and Emmanuel Schaan and Benjamin L. Schmitt and Neelima Sehgal and Shabbir Shaikh and Blake D. Sherwin and Carlos Sierra and Jon Sievers and Cristóbal Sifón and Sara Simon and Rita Sonka and David N. Spergel and Suzanne T. Staggs and Emilie Storer and Eric R. Switzer and Niklas Tampier and Robert Thornton and Hy Trac and Jesse Treu and Carole Tucker and Joel Ullom and Leila R. Vale and Alexander Van Engelen and Jeff Van Lanen and Cristian Vargas and Eve M. Vavagiakis and Kasey Wagoner and Yuhan Wang and Lukas Wenzl and Edward J. Wollack and Zhilei Xu and Fernando Zago and Kaiwen Zheng},
   doi = {10.1103/PhysRevD.109.063530},
   issn = {2470-0010},
   issue = {6},
   journal = {Physical Review D},
   month = {3},
   pages = {063530},
   title = {Atacama Cosmology Telescope: High-resolution component-separated maps across one third of the sky},
   volume = {109},
   url = {https://link.aps.org/doi/10.1103/PhysRevD.109.063530},
   year = {2024}
}

@article{DeBernardis2017DetectionTelescope,
    title = {{Detection of the pairwise kinematic Sunyaev-Zel'dovich effect with BOSS DR11 and the Atacama Cosmology Telescope}},
    year = {2017},
    journal = {Journal of Cosmology and Astroparticle Physics},
    author = {De Bernardis, F. and Aiola, S. and Vavagiakis, E.M. and Battaglia, N. and Niemack, M.D. and Beall, J. and Becker, D.T. and Bond, J.R. and Calabrese, E. and Cho, H. and Coughlin, K. and Datta, R. and Devlin, M. and Dunkley, J. and Dunner, R. and Ferraro, S. and Fox, A. and Gallardo, P.A. and Halpern, M. and Hand, N. and Hasselfield, M. and Henderson, S.W. and Hill, J.C. and Hilton, G.C. and Hilton, M. and Hincks, A.D. and Hlozek, R. and Hubmayr, J. and Huffenberger, K. and Hughes, J.P. and Irwin, K.D. and Koopman, B.J. and Kosowsky, A. and Li, D. and Louis, T. and Lungu, M. and Madhavacheril, M.S. and Maurin, L. and McMahon, J. and Moodley, K. and Naess, S. and Nati, F. and Newburgh, L. and Nibarger, J.P. and Page, L.A. and Partridge, B. and Schaan, E. and Schmitt, B. L. and Sehgal, N. and Sievers, J. and Simon, S.M. and Spergel, D.N. and Staggs, S.T. and Stevens, J.R. and Thornton, R.J. and Engelen, A. van and Lanen, J. Van and Wollack, E.J.},
    number = {03},
    month = {3},
    pages = {008--008},
    volume = {2017},
    publisher = {Institute of Physics Publishing},
    url = {https://iopscience.iop.org/article/10.1088/1475-7516/2017/03/008},
    doi = {10.1088/1475-7516/2017/03/008},
    issn = {1475-7516}
}

@article{DESICollaboration2025DataInstrument,
   author = {{DESI Collaboration} and M. Abdul-Karim and A. G. Adame and D. Aguado and J. Aguilar and S. Ahlen and S. Alam and G. Aldering and D. M. Alexander and R. Alfarsy and L. Allen and C. Allende Prieto and O. Alves and A. Anand and U. Andrade and E. Armengaud and S. Avila and A. Aviles and H. Awan and S. Bailey and A. Baleato Lizancos and O. Ballester and A. Bault and J. Bautista and S. BenZvi and L. Beraldo e Silva and J. R. Bermejo-Climent and F. Beutler and D. Bianchi and C. Blake and R. Blum and A. S. Bolton and M. Bonici and S. Brieden and A. Brodzeller and D. Brooks and E. Buckley-Geer and E. Burtin and R. Canning and A. Carnero Rosell and A. Carr and P. Carrilho and L. Casas and F. J. Castander and R. Cereskaite and J. L. Cervantes-Cota and E. Chaussidon and J. Chaves-Montero and S. Chen and X. Chen and T. Claybaugh and S. Cole and A. P. Cooper and M. -C. Cousinou and A. Cuceu and T. M. Davis and K. S. Dawson and R. de Belsunce and R. de la Cruz and A. de la Macorra and A. de Mattia and N. Deiosso and J. Della Costa and R. Demina and U. Demirbozan and J. DeRose and A. Dey and B. Dey and J. Ding and Z. Ding and P. Doel and K. Douglass and M. Dowicz and H. Ebina and J. Edelstein and D. J. Eisenstein and W. Elbers and N. Emas and S. Escoffier and P. Fagrelius and X. Fan and K. Fanning and V. A. Fawcett and E. Fernández-García and S. Ferraro and N. Findlay and A. Font-Ribera and J. E. Forero-Romero and D. Forero-Sánchez and C. S. Frenk and B. T. Gänsicke and L. Galbany and J. García-Bellido and C. Garcia-Quintero and L. H. Garrison and E. Gaztañaga and H. Gil-Marín and O. Y. Gnedin and S. Gontcho A Gontcho and A. X. Gonzalez-Morales and V. Gonzalez-Perez and C. Gordon and O. Graur and D. Green and D. Gruen and R. Gsponer and C. Guandalin and G. Gutierrez and J. Guy and C. Hahn and J. J. Han and J. Han and S. He and H. K. Herrera-Alcantar and K. Honscheid and J. Hou and C. Howlett and D. Huterer and V. Iršič and M. Ishak and A. Jacques and J. Jimenez and Y. P. Jing and B. Joachimi and S. Joudaki and R. Joyce and E. Jullo and S. Juneau and N. G. Karaçaylı and T. Karim and R. Kehoe and S. Kent and A. Khederlarian and D. Kirkby and T. Kisner and F. -S. Kitaura and N. Kizhuprakkat and H. Kong and S. E. Koposov and A. Kremin and A. Krolewski and O. Lahav and Y. Lai and C. Lamman and T. -W. Lan and M. Landriau and D. Lang and J. U. Lange and J. Lasker and J. M. Le Goff and L. Le Guillou and A. Leauthaud and M. E. Levi and S. Li and T. S. Li and K. Lodha and M. Lokken and Y. Luo and C. Magneville and M. Manera and C. J. Manser and D. Margala and P. Martini and M. Maus and J. McCullough and P. McDonald and G. E. Medina and L. Medina-Varela and A. Meisner and J. Mena-Fernández and A. Menegas and M. Mezcua and R. Miquel and P. Montero-Camacho and J. Moon and J. Moustakas and A. Muñoz-Gutiérrez and D. Muñoz-Santos and A. D. Myers and J. Myles and S. Nadathur and J. Najita and L. Napolitano and J. A. Newman and F. Nikakhtar and R. Nikutta and G. Niz and H. E. Noriega and N. Padmanabhan and E. Paillas and N. Palanque-Delabrouille and A. Palmese and J. Pan and Z. Pan and D. Parkinson and J. Peacock and W. J. Percival and A. Pérez-Fernández and I. Pérez-Ràfols and P. Peterson and J. Piat and M. M. Pieri and M. Pinon and C. Poppett and A. Porredon and F. Prada and R. Pucha and F. Qin and D. Rabinowitz and A. Raichoor and C. Ramírez-Pérez and S. Ramirez-Solano and M. Rashkovetskyi and C. Ravoux and A. H. Riley and A. Rocher and C. Rockosi and J. Rohlf and A. J. Ross and G. Rossi and R. Ruggeri and V. Ruhlmann-Kleider and C. G. Sabiu and K. Said and A. Saintonge and L. Samushia and E. Sanchez and N. Sanders and C. Saulder and E. F. Schlafly and D. Schlegel and D. Scholte and M. Schubnell and H. Seo and A. Shafieloo and R. Sharples and J. Silber and M. Siudek and A. Smith and D. Sprayberry and J. Suárez-Pérez and J. Swanson and T. Tan and G. Tarlé and P. Taylor and G. Thomas and R. Tojeiro and R. J. Turner and W. Turner and L. A. Ureña-López and R. Vaisakh and M. Valluri and M. Vargas-Magaña and L. Verde and M. Walther and B. Wang and M. S. Wang and W. Wang and B. A. Weaver and N. Weaverdyck and R. H. Wechsler and M. White and M. Wolfson and J. Yang and C. Yèche and S. Youles and J. Yu and S. Yuan and E. A. Zaborowski and P. Zarrouk and H. Zhang and C. Zhao and R. Zhao and Z. Zheng and R. Zhou and H. Zou and S. Zou and Y. Zu},
   month = {3},
   title = {Data Release 1 of the Dark Energy Spectroscopic Instrument},
   journal = {ArXiv e-prints},
   archivePrefix = "arXiv",
   eprint = {2503.14745},
   url = {http://arxiv.org/abs/2503.14745},
   year = {2025}
}

@article{Ding2024MiscenteringCounterparts,
    title = {{Miscentering of Optical Galaxy Clusters Based on Sunyaev-Zeldovich Counterparts}},
    year = {2024},
    journal = {MNRAS},
    author = {Ding, Jupiter and Dalal, Roohi and Sunayama, Tomomi and Strauss, Michael A. and Oguri, Masamune and Okabe, Nobuhiro and Hilton, Matt and Monteiro-Oliveira, Rogério and Sif{\'{o}}n, Cristóbal and Staggs, Suzanne T.},
    number = {1},
    month = {11},
    pages = {572--591},
    volume = {536},
    url = {http://arxiv.org/abs/2411.12120 http://dx.doi.org/10.1093/mnras/stae2601},
    doi = {10.1093/mnras/stae2601},
    arxivId = {2411.12120}
}

@article{Dolag2009SubstructuresSimulations,
    title = {{Substructures in hydrodynamical cluster simulations}},
    year = {2009},
    journal = {Monthly Notices of the Royal Astronomical Society},
    author = {Dolag, K. and Borgani, S. and Murante, G. and Springel, V.},
    number = {2},
    month = {10},
    pages = {497--514},
    volume = {399},
    publisher = {Blackwell Publishing Ltd},
    url = {https://academic.oup.com/mnras/article-lookup/doi/10.1111/j.1365-2966.2009.15034.x},
    doi = {10.1111/j.1365-2966.2009.15034.x},
    issn = {00358711}
}

@article{Dolag2005ThePlanck,
    title = {{The imprints of local superclusters on the Sunyaev-Zel'dovich signals and their detectability with Planck}},
    year = {2005},
    journal = {Monthly Notices of the Royal Astronomical Society},
    author = {Dolag, K. and Hansen, F. K. and Roncarelli, M. and Moscardini, L.},
    number = {1},
    month = {10},
    pages = {29--39},
    volume = {363},
    publisher = {Blackwell Publishing Ltd},
    url = {https://ui.adsabs.harvard.edu/abs/2005MNRAS.363...29D/abstract},
    doi = {10.1111/j.1365-2966.2005.09452.x},
    issn = {00358711},
    arxivId = {astro-ph/0505258}
}

@article{Dolag2016,
    title = {{SZ effects in the Magneticum Pathfinder simulation: comparison with the Planck, SPT, and ACT results}},
    year = {2016},
    journal = {Monthly Notices of the Royal Astronomical Society},
    author = {Dolag, K. and Komatsu, E. and Sunyaev, R.},
    number = {2},
    month = {12},
    pages = {1797--1811},
    volume = {463},
    publisher = {Oxford Academic},
    url = {https://academic.oup.com/mnras/article/463/2/1797/2892389},
    doi = {10.1093/MNRAS/STW2035},
    issn = {0035-8711},
    arxivId = {1509.05134}
}

@article{Dolag2025EncyclopediaDay,
    title = {{Encyclopedia Magneticum: Scaling Relations from Cosmic Dawn to Present Day}},
    year = {2025},
    author = {Dolag, Klaus and Remus, Rhea-Silvia and Valenzuela, Lucas M. and Kimmig, Lucas C. and Seidel, Benjamin and Fortune, Silvio and Stoiber, Johannes and Ivleva, Anna and Hoffmann, Tadziu and Biffi, Veronica and Marini, Ilaria and Popesso, Paola and Vladutescu-Zopp, Stephan},
    month = {4},
    url = {https://arxiv.org/abs/2504.01061v1},
    arxivId = {2504.01061},
    journal = {ArXiv e-prints},
    archivePrefix = "arXiv",
    eprint = {2504.01061}
}

@article{Ferreira1999Streaming,
    title = {{Streaming Velocities as a Dynamical Estimator of {$\Omega$}}},
    year = {1999},
    journal = {The Astrophysical Journal},
    author = {Ferreira, P. G. and Juszkiewicz, R. and Feldman, H. A. and Davis, M. and Jaffe, A. H.},
    number = {1},
    month = {4},
    pages = {L1-L4},
    volume = {515},
    publisher = {American Astronomical Society},
    url = {https://ui.adsabs.harvard.edu/abs/1999ApJ...515L...1F/abstract},
    doi = {10.1086/311959},
    issn = {0004637X},
    arxivId = {astro-ph/9812456}
}

@article{Gallardo2019OptimizingTelescope,
    title = {{Optimizing Future Cmb Observatories and Measuring Galaxy Cluster Motions With the Atacama Cosmology Telescope}},
    year = {2019},
    author = {Gallardo, P. A.},
    number = {December},
    note = {https://doi.org/10.7298/kj4t-8e29},
    url = {https://ecommons.cornell.edu/handle/1813/70103}
}

@article{Gallardo2025,
doi = {10.3847/2515-5172/ae1587},
url = {https://doi.org/10.3847/2515-5172/ae1587},
year = {2025},
month = {oct},
publisher = {The American Astronomical Society},
volume = {9},
number = {10},
pages = {284},
author = {Gallardo, Patricio A. and Gong, Yulin and Hadzhiyska, Boryana and Hsu, Yun-Hsin},
title = {Iskay2: Signal Extraction of the Kinematic Sunyaev–Zel'dovich Effect Through the Pairwise Estimator: Pipeline and Validation},
journal = {Research Notes of the AAS}
}

@ARTICLE{Gong2025,
       author = {{Gong}, Yulin and {Gallardo}, Patricio A. and {Bean}, Rachel and {Moore}, Jenna and {Vavagiakis}, Eve M. and {Battaglia}, Nicholas and {Hadzhiyska}, Boryana and {Hsu}, Yun-Hsin and {Aguilar}, Jessica Nicole and {Ahlen}, Steven and {Bianchi}, Davide and {Brooks}, David and {Claybaugh}, Todd and {Canning}, Rebecca and {Devlin}, Mark and {Doel}, Peter and {de la Macorra}, Axel and {Ferraro}, Simone and {Font-Ribera}, Andreu and {Forero-Romero}, Jaime E. and {Gazta{\~n}aga}, Enrique and {Gutierrez}, Gaston and {Gontcho}, Satya Gontcho A and {Guy}, Julien and {Honscheid}, Klaus and {Howlett}, Cullan and {Liu}, R. Henry and {Ishak}, Mustapha and {Joyce}, Dick and {Kremin}, Anthony and {Lamman}, Claire and {Levi}, Michael and {Landriau}, Martin and {Manera}, Marc and {Meisner}, Aaron and {Miquel}, Ramon and {Niemack}, Michael D. and {Nadathur}, Seshadri and {Percival}, Will and {Prada}, Francisco and {Rossi}, Graziano and {Ried Guachalla}, Bernardita and {Sanchez}, Eusebio and {Seo}, Hee-Jong and {Sprayberry}, David and {Schlegel}, David and {Sif{\'o}n}, Crist{\'o}bal and {Schubnell}, Michael and {Silber}, Joseph Harry and {Tarl{\'e}}, Gregory and {Weaver}, Benjamin Alan and {Zhou}, Rongpu and {Zou}, Hu},
        title = "{Detection of the Pairwise Kinematic Sunyaev-Zel'dovich Effect and Pairwise Velocity with DESI DR1 Galaxies and ACT DR6 and Planck CMB Data}",
      journal = {arXiv e-prints},
         year = 2025,
        month = nov,
          eid = {arXiv:2511.23417},
        pages = {arXiv:2511.23417},
          doi = {10.48550/arXiv.2511.23417},
archivePrefix = {arXiv},
       eprint = {2511.23417},
 primaryClass = {astro-ph.CO},
       adsurl = {https://ui.adsabs.harvard.edu/abs/2025arXiv251123417G}
}

@article{Hadzhiyska2024EvidenceGalaxies,
    title = {{Evidence for large baryonic feedback at low and intermediate redshifts from kinematic Sunyaev-Zel'dovich observations with ACT and DESI photometric galaxies}},
    year = {2024},
    author = {Hadzhiyska, B. and Ferraro, S. and Guachalla, B. Ried and Schaan, E. and Aguilar, J. and Battaglia, N. and Bond, J. R. and Brooks, D. and Calabrese, E. and Choi, S. K. and Claybaugh, T. and Coulton, W. R. and Dawson, K. and Devlin, M. and Dey, B. and Doel, P. and Duivenvoorden, A. J. and Dunkley, J. and Farren, G. S. and Font-Ribera, A. and Forero-Romero, J. E. and Gallardo, P. A. and Gazta{\~{n}}aga, E. and Gontcho, S. Gontcho and Gralla, M. and Guillou, L. Le and Gutierrez, G. and Guy, J. and Hill, J. C. and Hlo{\v{z}}ek, R. and Honscheid, K. and Juneau, S. and Kisner, T. and Kremin, A. and Landriau, M. and Liu, R. H. and Louis, T. and MacCrann, N. and de Macorra, A. and Madhavacheril, M. and Manera, M. and Meisner, A. and Miquel, R. and Moodley, K. and Moustakas, J. and Mroczkowski, T. and Naess, S. and Newman, J. and Niemack, M. D. and Niz, G. and Page, L. and Palanque-Delabrouille, N. and Partridge, B. and Percival, W. J. and Prada, F. and Qu, F. J. and Rossi, G. and Sanchez, E. and Schlegel, D. and Schubnell, M. and Sehgal, N. and Seo, H. and Sif{\'{o}}n, C. and Spergel, D. and Sprayberry, D. and Staggs, S. and Tarl{\'{e}}, G. and Vargas, C. and Vavagiakis, E. M. and Weaver, B. A. and Wollack, E. J. and Zhou, R. and Zou, H.},
    month = {7},
    pages = {},
    volume = {},
    url = {http://arxiv.org/abs/2407.07152},
    journal = {ArXiv e-prints},
    archivePrefix = "arXiv",
    eprint = {2407.07152},
    arxivId = {2407.07152}
}

@ARTICLE{Hadzhiyska2025,
       author = {{Hadzhiyska}, B. and {Gong}, Y. and {Hsu}, Y. and {Gallardo}, P.~A. and {Aguilar}, J. and {Ahlen}, S. and {Alonso}, D. and {Bean}, R. and {Bianchi}, D. and {Brooks}, D. and {Castander}, F.~J. and {Claybaugh}, T. and {Cole}, S. and {Cuceu}, A. and {de la Macorra}, A. and {Dey}, Arjun and {Ferraro}, S. and {Font-Ribera}, A. and {Forero-Romero}, J.~E. and {Gontcho}, S. Gontcho A and {Gutierrez}, G. and {Guy}, J. and {Herrera-Alcantar}, H.~K. and {Howlett}, C. and {Huterer}, D. and {Ishak}, M. and {Joyce}, R. and {Kisner}, T. and {Kremin}, A. and {Landriau}, M. and {Le Guillou}, L. and {Levi}, M.~E. and {Manera}, M. and {Meisner}, A. and {Miquel}, R. and {Moodley}, K. and {Mroczkowski}, T. and {Nadathur}, S. and {Palanque-Delabrouille}, N. and {Percival}, W.~J. and {Prada}, F. and {Qu}, F.~J. and {Perez-Rafols}, I. and {Ried Guachalla}, B. and {Rossi}, G. and {Sanchez}, E. and {Schaan}, E. and {Schlegel}, D. and {Schubnell}, M. and {Seo}, H. and {Sifon}, C. and {Silber}, J. and {Sprayberry}, D. and {Tarle}, G. and {Vavagiakis}, E.~M. and {Weaver}, B.~A. and {Zhou}, R. and {Zou}, H.},
        title = "{Probing cosmic velocities with the pairwise kinematic Sunyaev-Zel'dovich signal in DESI Bright Galaxy Sample DR1 and ACT DR6}",
      journal = {arXiv e-prints},
         year = 2025,
        month = oct,
          eid = {arXiv:2510.14135},
        pages = {arXiv:2510.14135},
          doi = {10.48550/arXiv.2510.14135},
archivePrefix = {arXiv},
       eprint = {2510.14135},
 primaryClass = {astro-ph.CO},
       adsurl = {https://ui.adsabs.harvard.edu/abs/2025arXiv251014135H}
}

@article{Hand2012EvidenceEffect,
    title = {{Evidence of Galaxy Cluster Motions with the Kinematic Sunyaev-Zel’dovich Effect}},
    year = {2012},
    journal = {Physical Review Letters},
    author = {Hand, Nick and Addison, Graeme E. and Aubourg, Eric and Battaglia, Nick and Battistelli, Elia S. and Bizyaev, Dmitry and Bond, J. Richard and Brewington, Howard and Brinkmann, Jon and Brown, Benjamin R. and Das, Sudeep and Dawson, Kyle S. and Devlin, Mark J. and Dunkley, Joanna and Dunner, Rolando and Eisenstein, Daniel J. and Fowler, Joseph W. and Gralla, Megan B. and Hajian, Amir and Halpern, Mark and Hilton, Matt and Hincks, Adam D. and Hlozek, Renée and Hughes, John P. and Infante, Leopoldo and Irwin, Kent D. and Kosowsky, Arthur and Lin, Yen-Ting and Malanushenko, Elena and Malanushenko, Viktor and Marriage, Tobias A. and Marsden, Danica and Menanteau, Felipe and Moodley, Kavilan and Niemack, Michael D. and Nolta, Michael R. and Oravetz, Daniel and Page, Lyman A. and Palanque-Delabrouille, Nathalie and Pan, Kaike and Reese, Erik D. and Schlegel, David J. and Schneider, Donald P. and Sehgal, Neelima and Shelden, Alaina and Sievers, Jon and Sif{\'{o}}n, Cristóbal and Simmons, Audrey and Snedden, Stephanie and Spergel, David N. and Staggs, Suzanne T. and Swetz, Daniel S. and Switzer, Eric R. and Trac, Hy and Weaver, Benjamin A. and Wollack, Edward J. and Yeche, Christophe and Zunckel, Caroline},
    number = {4},
    month = {7},
    pages = {041101},
    volume = {109},
    url = {https://link.aps.org/doi/10.1103/PhysRevLett.109.041101},
    doi = {10.1103/PhysRevLett.109.041101},
    issn = {0031-9007}
}

@article{Hirschmann2014CosmologicalDownsizing,
    title = {{Cosmological simulations of black hole growth: AGN luminosities and downsizing}},
    year = {2014},
    journal = {Monthly Notices of the Royal Astronomical Society},
    author = {Hirschmann, M. and Dolag, K. and Saro, A. and Bachmann, L. and Borgani, S. and Burkert, A.},
    number = {3},
    month = {6},
    pages = {2304--2324},
    volume = {442},
    publisher = {Oxford Academic},
    url = {https://academic.oup.com/mnras/article-lookup/doi/10.1093/mnras/stu1023},
    doi = {10.1093/mnras/stu1023},
    issn = {0035-8711}
}

@article{Kluge2024TheSurvey,
    title = {{The SRG/eROSITA All-Sky Survey}},
    year = {2024},
    journal = {Astronomy {\&} Astrophysics},
    author = {Kluge, M. and Comparat, J. and Liu, A. and Balzer, F. and Bulbul, E. and Ider Chitham, J. and Ghirardini, V. and Garrel, C. and Bahar, Y. E. and Artis, E. and Bender, R. and Clerc, N. and Dwelly, T. and Fabricius, M. H. and Grandis, S. and Hern{\'{a}}ndez-Lang, D. and Hill, G. J. and Joshi, J. and Lamer, G. and Merloni, A. and Nandra, K. and Pacaud, F. and Predehl, P. and Ramos-Ceja, M. E. and Reiprich, T. H. and Salvato, M. and Sanders, J. S. and Schrabback, T. and Seppi, R. and Zelmer, S. and Zenteno, A. and Zhang, X.},
    month = {8},
    pages = {A210},
    volume = {688},
    url = {https://www.aanda.org/10.1051/0004-6361/202349031},
    doi = {10.1051/0004-6361/202349031},
    issn = {0004-6361}
}

@article{Komatsu2011SEVEN-YEARINTERPRETATION,
    title = {{SEVEN-YEAR <i>WILKINSON MICROWAVE ANISOTROPY PROBE</i> ( <i>WMAP</i> ) OBSERVATIONS: COSMOLOGICAL INTERPRETATION}},
    year = {2011},
    journal = {The Astrophysical Journal Supplement Series},
    author = {Komatsu, E. and Smith, K. M. and Dunkley, J. and Bennett, C. L. and Gold, B. and Hinshaw, G. and Jarosik, N. and Larson, D. and Nolta, M. R. and Page, L. and Spergel, D. N. and Halpern, M. and Hill, R. S. and Kogut, A. and Limon, M. and Meyer, S. S. and Odegard, N. and Tucker, G. S. and Weiland, J. L. and Wollack, E. and Wright, E. L.},
    number = {2},
    month = {2},
    pages = {18},
    volume = {192},
    url = {https://iopscience.iop.org/article/10.1088/0067-0049/192/2/18},
    doi = {10.1088/0067-0049/192/2/18},
    issn = {0067-0049}
}

@article{Levi2013The2013,
    title = {{The DESI Experiment, a whitepaper for Snowmass 2013}},
    year = {2013},
    journal = {arXiv},
    author = {Levi, Michael and Bebek, Chris and Beers, Timothy and Blum, Robert and Cahn, Robert and Eisenstein, Daniel and Flaugher, Brenna and Honscheid, Klaus and Kron, Richard and Lahav, Ofer and McDonald, Patrick and Roe, Natalie and Schlegel, David and collaboration, representing the DESI and Levi, Michael and Bebek, Chris and Beers, Timothy and Blum, Robert and Cahn, Robert and Eisenstein, Daniel and Flaugher, Brenna and Honscheid, Klaus and Kron, Richard and Lahav, Ofer and McDonald, Patrick and Roe, Natalie and Schlegel, David and collaboration, representing the DESI},
    month = {8},
    pages = {arXiv:1308.0847},
    url = {https://ui.adsabs.harvard.edu/abs/2013arXiv1308.0847L/abstract},
    arxivId = {1308.0847}
}

@article{Li2024DetectionSpace,
    title = {{Detection of Pairwise Kinetic Sunyaev–Zel’dovich Effect with DESI Galaxy Groups and Planck in Fourier Space}},
    year = {2024},
    journal = {The Astrophysical Journal Supplement Series},
    author = {Li, Shaohong and Zheng, Yi and Chen, Ziyang and Xu, Haojie and Yang, Xiaohu},
    number = {1},
    month = {3},
    pages = {30},
    volume = {271},
    publisher = {IOP Publishing},
    url = {https://iopscience.iop.org/article/10.3847/1538-4365/ad1bd8 https://iopscience.iop.org/article/10.3847/1538-4365/ad1bd8/meta},
    doi = {10.3847/1538-4365/AD1BD8},
    issn = {0067-0049}
}

@article{Marini2025DetectingSurveys,
    title = {{Detecting clusters and groups of galaxies populating the local Universe in large optical spectroscopic surveys}},
    year = {2025},
    journal = {Astronomy {\&} Astrophysics},
    author = {Marini, I. and Popesso, P. and Dolag, K. and Bravo, M. and Robotham, A. and Tempel, E. and Li, Q. and Yang, X. and Csizi, B. and Behroozi, P. and Biffi, V. and Biviano, A. and Lamer, G. and Malavasi, N. and Mazengo, D. and Toptun, V.},
    month = {2},
    pages = {A207},
    volume = {694},
    publisher = {EDP Sciences},
    url = {https://www.aanda.org/10.1051/0004-6361/202452028},
    doi = {10.1051/0004-6361/202452028},
    issn = {0004-6361}
}

@article{Marini2024DetectingEra,
    title = {{Detecting galaxy groups populating the local Universe in the eROSITA era}},
    year = {2024},
    journal = {Astronomy {\&} Astrophysics},
    author = {Marini, I. and Popesso, P. and Lamer, G. and Dolag, K. and Biffi, V. and Vladutescu-Zopp, S. and Dev, A. and Toptun, V. and Bulbul, E. and Comparat, J. and Malavasi, N. and Merloni, A. and Mroczkowski, T. and Ponti, G. and Seppi, R. and Shreeram, S. and Zhang, Y.},
    month = {9},
    pages = {A7},
    volume = {689},
    publisher = {EDP Sciences},
    url = {https://www.aanda.org/10.1051/0004-6361/202450442},
    doi = {10.1051/0004-6361/202450442},
    issn = {0004-6361}
}

@article{McCarthy2024FLAMINGO:Structure,
    title = {{FLAMINGO: combining kinetic SZ effect and galaxy-galaxy lensing measurements to gauge the impact of feedback on large-scale structure}},
    year = {2024},
    author = {McCarthy, Ian G. and Amon, Alexandra and Schaye, Joop and Schaan, Emmanuel and Angulo, Raul E. and Salcido, Jaime and Schaller, Matthieu and Bigwood, Leah and Elbers, Willem and Kugel, Roi and Helly, John C. and Moreno, Victor J. Forouhar and Frenk, Carlos S. and McGibbon, Robert J. and Ondaro-Mallea, Lurdes and van Daalen, Marcel P.},
    month = {10},
    pages = {},
    volume = {},
    url = {http://arxiv.org/abs/2410.19905},
    arxivId = {2410.19905},
    journal = {ArXiv e-prints},
    archivePrefix = "arXiv",
    eprint = {2410.19905}
}

@article{Mueller2014ConstraintsEffect,
    title = {{Constraints on gravity and dark energy from the pairwise kinematic Sunyaev-Zeldovich effect}},
    year = {2014},
    journal = {The Astrophysical Journal},
    author = {Mueller, Eva-Maria and de Bernardis, Francesco and Bean, Rachel and Niemack, Michael},
    number = {1},
    month = {8},
    pages = {47},
    volume = {808},
    publisher = {Institute of Physics Publishing},
    url = {http://arxiv.org/abs/1408.6248 http://dx.doi.org/10.1088/0004-637X/808/1/47},
    doi = {10.1088/0004-637X/808/1/47},
    arxivId = {1408.6248}
}

@article{Mueller,
    title = {{Constraints on massive neutrinos from the pairwise kinematic Sunyaev-Zel’dovich effect}},
    year = {2015},
    journal = {Physical Review D},
    author = {Mueller, Eva-Maria and de Bernardis, Francesco and Bean, Rachel and Niemack, Michael D.},
    number = {6},
    month = {9},
    pages = {063501},
    volume = {92},
    url = {https://link.aps.org/doi/10.1103/PhysRevD.92.063501},
    doi = {10.1103/PhysRevD.92.063501},
    issn = {1550-7998}
}

@article{Myles2021SpectroscopicCatalogue,
    title = {{Spectroscopic quantification of projection effects in the SDSS redMaPPer galaxy cluster catalogue}},
    year = {2021},
    journal = {Monthly Notices of the Royal Astronomical Society},
    author = {Myles, J and Gruen, D and Mantz, A B and Allen, S W and Morris, R G and Rykoff, E and Costanzi, M and To, C and DeRose, J and Wechsler, R H and Rozo, E and Jeltema, T and Carrasco, E R and Kremin, A and Kron, R},
    number = {1},
    month = {5},
    pages = {33--44},
    volume = {505},
    publisher = {arXiv},
    url = {https://academic.oup.com/mnras/article/505/1/33/6262362},
    doi = {10.1093/mnras/stab1243},
    issn = {0035-8711}
}

@article{Naess2025,
   author = {Sigurd Naess and Yilun Guan and Adriaan J. Duivenvoorden and Matthew Hasselfield and Yuhan Wang and Irene Abril-Cabezas and Graeme E. Addison and Peter A. R. Ade and Simone Aiola and Tommy Alford and David Alonso and Mandana Amiri and Rui An and Zachary Atkins and Jason E. Austermann and Eleonora Barbavara and Nicholas Battaglia and Elia Stefano Battistelli and James A. Beall and Rachel Bean and Ali Beheshti and Benjamin Beringue and Tanay Bhandarkar and Emily Biermann and Boris Bolliet and J Richard Bond and Erminia Calabrese and Valentina Capalbo and Felipe Carrero and Stephen Chen and Grace Chesmore and Hsiao-mei Cho and Steve K. Choi and Susan E. Clark and Rodrigo Cordova Rosado and Nicholas F. Cothard and Kevin Coughlin and William Coulton and Devin Crichton and Kevin T. Crowley and Mark J. Devlin and Simon Dicker and Cody J. Duell and Shannon M. Duff and Jo Dunkley and Rolando Dunner and Carmen Embil Villagra and Max Fankhanel and Gerrit S. Farren and Simone Ferraro and Allen Foster and Rodrigo Freundt and Brittany Fuzia and Patricio A. Gallardo and Xavier Garrido and Serena Giardiello and Ajay Gill and Jahmour Givans and Vera Gluscevic and Joseph E. Golec and Yulin Gong and Mark Halpern and Ian Harrison and Erin Healy and Shawn Henderson and Brandon Hensley and Carlos Hervías-Caimapo and J. Colin Hill and Gene C. Hilton and Matt Hilton and Adam D. Hincks and Renée Hložek and Shuay-Pwu Patty Ho and John Hood and Erika Hornecker and Zachary B. Huber and Johannes Hubmayr and Kevin M. Huffenberger and John P. Hughes and Margaret Ikape and Kent Irwin and Giovanni Isopi and Hidde T. Jense and Neha Joshi and Ben Keller and Joshua Kim and Kenda Knowles and Brian J. Koopman and Arthur Kosowsky and Darby Kramer and Aleksandra Kusiak and Adrien La Posta and Alex Laguë and Victoria Lakey and Eunseong Lee and Yaqiong Li and Zack Li and Michele Limon and Martine Lokken and Thibaut Louis and Marius Lungu and Niall MacCrann and Amanda MacInnis and Mathew S. Madhavacheril and Diego Maldonado and Felipe Maldonado and Maya Mallaby-Kay and Gabriela A. Marques and Joshiwa van Marrewijk and Fiona McCarthy and Jeff McMahon and Yogesh Mehta and Felipe Menanteau and Kavilan Moodley and Thomas W. Morris and Tony Mroczkowski and Toshiya Namikawa and Federico Nati and Simran K. Nerval and Laura Newburgh and Andrina Nicola and Michael D. Niemack and Michael R. Nolta and John Orlowski-Scherer and Lyman A. Page and Shivam Pandey and Bruce Partridge and Karen Perez Sarmiento and Heather Prince and Roberto Puddu and Frank J. Qu and Rodrigo Quiroga and Damien C. Ragavan and Bernardita Ried Guachalla and Keir K. Rogers and Felipe Rojas and Tai Sakuma and Emmanuel Schaan and Benjamin L. Schmitt and Neelima Sehgal and Shabbir Shaikh and Blake D. Sherwin and Carlos Sierra and Jon Sievers and Cristóbal Sifón and Sara Simon and Rita Sonka and David N. Spergel and Suzanne T. Staggs and Emilie Storer and Kristen Surrao and Eric R. Switzer and Niklas Tampier and Robert Thornton and Hy Trac and Carole Tucker and Joel Ullom and Leila R. Vale and Alexander Van Engelen and Jeff Van Lanen and Cristian Vargas and Eve M. Vavagiakis and Kasey Wagoner and Lukas Wenzl and Edward J. Wollack and Kaiwen Zheng},
   journal = {arXiv},
   month = {3},
   pages = {arXiv:2503.14451},
   title = {The Atacama Cosmology Telescope: DR6 Maps},
   url = {http://arxiv.org/abs/2503.14451},
   year = {2025}
}

@article{Orlowski-Scherer2021AtacamaACT,
    title = {{Atacama Cosmology Telescope measurements of a large sample of candidates from the Massive and Distant Clusters of WISE Survey: Sunyaev-Zeldovich effect confirmation of MaDCoWS candidates using ACT}},
    year = {2021},
    journal = {Astronomy and Astrophysics},
    author = {Orlowski-Scherer, John and Di Mascolo, Luca and Bhandarkar, Tanay and Manduca, Alex and Mroczkowski, Tony and Amodeo, Stefania and Battaglia, Nick and Brodwin, Mark and Choi, Steve K. and Devlin, Mark and Dicker, Simon and Dunkley, Jo and Gonzalez, Anthony H. and Han, Dongwon and Hilton, Matt and Huffenberger, Kevin and Hughes, John P. and MacInnis, Amanda and Knowles, Kenda and Koopman, Brian J. and Lowe, Ian and Moodley, Kavilan and Nati, Federico and Niemack, Michael D. and Page, Lyman A. and Partridge, Bruce and Romero, Charles and Salatino, Maria and Schillaci, Alessandro and Sehgal, Neelima and Sif{\'{o}}n, Cristóbal and Staggs, Suzanne and Stanford, Spencer A. and Thornton, Robert and Vavagiakis, Eve M. and Wollack, Edward J. and Xu, Zhilei and Zhu, Ningfeng},
    month = {9},
    pages = {A135},
    volume = {653},
    publisher = {EDP Sciences},
    url = {https://ui.adsabs.harvard.edu/abs/2021A&A...653A.135O/abstract},
    doi = {10.1051/0004-6361/202141200},
    issn = {14320746},
    arxivId = {2105.00068}
}

@ARTICLE{Osato2018,
       author = {{Osato}, Ken and {Nishimichi}, Takahiro and {Oguri}, Masamune and {Takada}, Masahiro and {Okumura}, Teppei},
        title = "{Strong orientation dependence of surface mass density profiles of dark haloes at large scales}",
      journal = {\mnras},
         year = 2018,
        month = jun,
       volume = {477},
       number = {2},
        pages = {2141-2153},
          doi = {10.1093/mnras/sty762},
archivePrefix = {arXiv},
       eprint = {1712.00094},
 primaryClass = {astro-ph.CO},
       adsurl = {https://ui.adsabs.harvard.edu/abs/2018MNRAS.477.2141O}
}

\begin{appendix}

\section{Matched filter}\label{appendix:MF} 

We employed the matched filter approach to combine the pairwise kSZ signal over a range of separations in a way that optimizes the S/N. Assume that the data vector $\bf{d}$ is a linear combination of a signal template $\mathbf{s}_\mathrm{m}(r)$ , amplitude $a_0$, and noise $\mathbf{n}_\mathrm{m}$,
\[ \mathbf{d} = a_0 \mathbf{s}_\mathrm{m}(r)+\mathbf{n}_\mathrm{m} \,,\]
We can find a linear transform that maximizes the S/N of our $a_0$ estimation\,(see e.g.\,\citealt{Zubeldia2021UnderstandingCosmology} for the detailed derivation), as
\[ \mathbf{T} = \frac{\mathbf{s}_\mathrm{m}^T \mathbf{C}_\mathrm{m}^{-1}}{\mathbf{s}_\mathrm{m}^T \mathbf{C}_\mathrm{m}^{-1} \mathbf{s}_\mathrm{m}} \]
where $\mathbf{C}_\mathrm{m}$ is the noise covariance matrix model. The optimized S/N is given as
\[ \hat{a}_0 = \frac{\mathbf{s}_\mathrm{m}^T \mathbf{C}_\mathrm{m}^{-1}\mathbf{d}}{\mathbf{s}_\mathrm{m}^T \mathbf{C}_\mathrm{m}^{-1} \mathbf{s}_\mathrm{m}}, \quad \sigma_a^2 = \frac{1}{\mathbf{s}_\mathrm{m}^T \mathbf{C}_\mathrm{m}^{-1} \mathbf{s}_\mathrm{m}}, \quad \frac{S}{N} = \frac{\hat{a}_0}{\sigma_a}\:.\]

We obtained the signal template $\mathbf{s}_\mathrm{m}$ and the covariance template $\mathbf{C}_\mathrm{m}$ empirically from the \texttt{Magneticum} $35\times35\,\mathrm{deg^2}$ light cone\,\citep{Soergel2018CosmologySimulations}. We fitted a smoothed broken power law function to the pairwise velocity profile of the $z=0.2-0.6$ clusters up to 150 cMpc and adopted it as $\bf{s_m}$. The potential redshift evolution was corrected (see Sect.\,\ref{subsec:kSZ and components}). The smoothing parameter $\Delta$ was set to 1:
\[
f(x) = A \left( \frac{x}{x_b} \right)^{-\alpha_1} 
\left\{ \frac{1}{2} \left[ 1 + \left( \frac{x}{x_b} \right)^{1/\Delta} \right] \right\}^{(\alpha_1 - \alpha_2) \Delta} \,.
\]
There were 23,\,177 clusters selected with $M_{vir}\geq 10^{13.5}\,h^{-1}M_{\odot}$, which is similar to the mass cut corresponding to $\lambda \geq 5$, which is $M_{200m} \geq 10^{13.4} h^{-1}M_{\odot}$. Although this selection differs slightly from our main sample due to the simulation resolution, the template is only used to obtain a higher S/N while summing up the profile, and the main result is not sensitive to the shape of the template.

\begin{figure}[h!]
\begin{minipage}[t]{1\columnwidth}
\includegraphics[width=\linewidth,height=0.34\textheight]{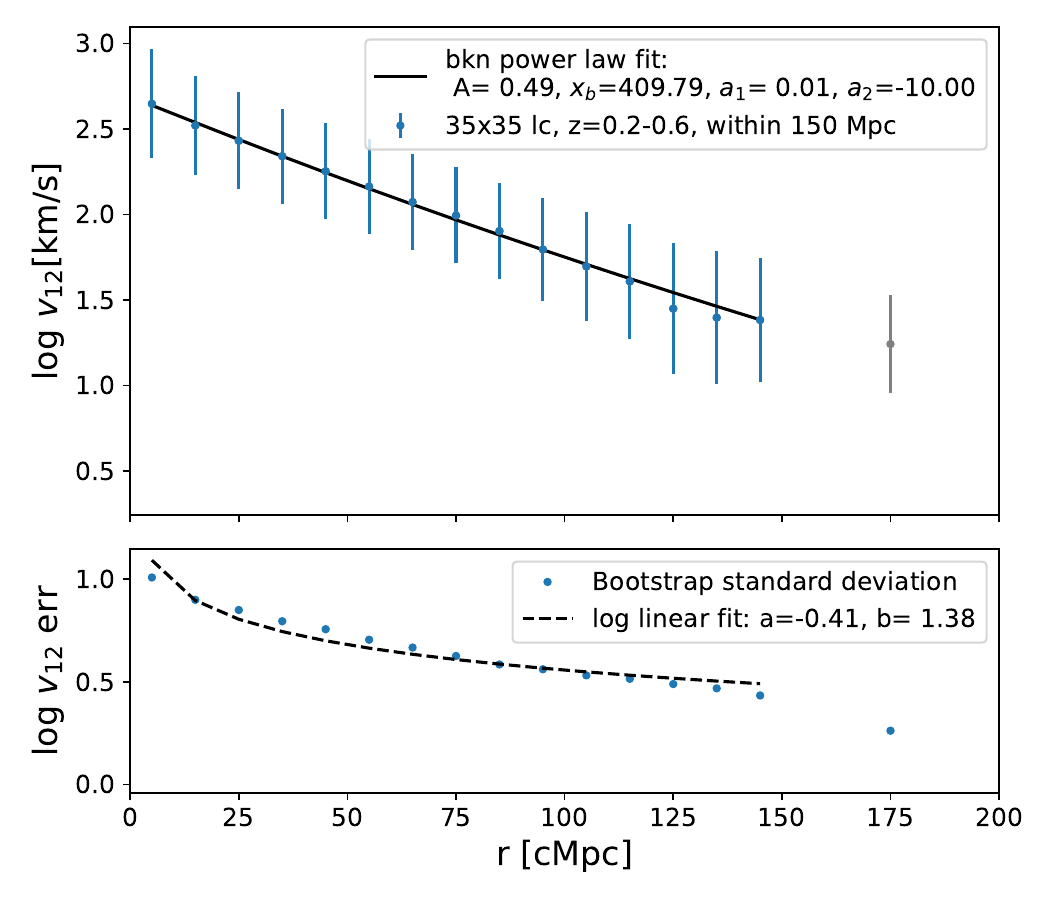}
\caption{Signal and covariance templates for the optimized filter, obtained from the pairwise velocity of clusters in the $35\times35\,\mathrm{deg^2}$ light cone. The fitting parameters for the empirical templates are presented in the plot. The x-axis indicates the comoving separation of the cluster pairs. Templates are fitted up to 150\,$\mathrm{cMpc}$.}
\label{fig:MF}
\end{minipage}
\hfill
\begin{minipage}[t]{1\columnwidth}
\includegraphics[width=\linewidth,height=0.22\textheight]{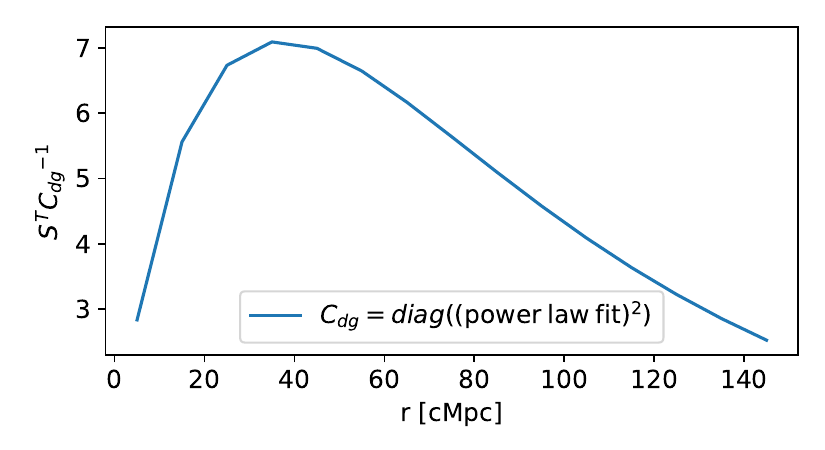}
\caption{Illustration of the shape of the numerator of the filter $\hat{a}_0$. The covariance $\mathbf{C}_\mathrm{m}$ is assumed to be diagonal and is presented as $\mathbf{C}_\mathrm{dg}$.}
\label{fig:MF_diag}
\end{minipage}
\end{figure}

The error model was obtained by fitting a log-linear relation, to the logarithm of the standard deviation of the 4000 times bootstrap realizations of the pairwise kSZ curve. We assumed the covariance is highly diagonal, which is reasonable within 150 cMpc\,(Fig.\,\ref{fig:MF}). The bias result is based on the ratio between two signals, so the normalization of the filter is irrelevant -- only the shape matters. If we ignore the normalization denominator, the numerator $\mathbf{T}\approx\mathbf{s}_\mathrm{m}^T \mathbf{C}_\mathrm{m}^{-1}$ weights the signal as shown in Fig.\,\ref{fig:MF_diag}. The shape of the template captures two critical features: there is little signal beyond 150 cMpc, and the uncertainties are large within 20 cMpc.

\section{Light-cone sky coordinates approximation}\label{appendix:lc_coord} 

\begin{figure}
\centering
    \includegraphics[width=0.5\textwidth]{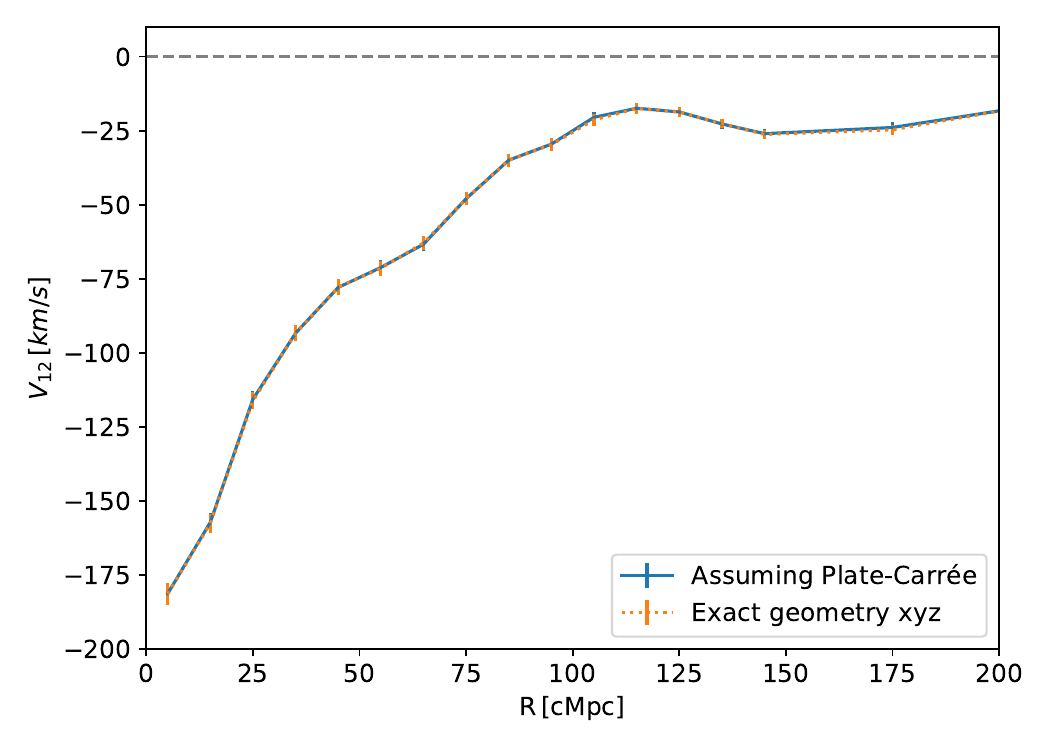}
    \caption{Comparison between the exact distance calculation and the approximate sky coordinates used in the mock catalog. Pairwise velocities computed from exact Cartesian\,$(X, Y, Z)$ coordinates\,(orange dotted line), and from the mock Plate-Carr\'{e}e coordinates\,(blue solid line) are fully consistent. Uncertainties are estimated by bootstrapping the cluster catalog 1000 times.}
    \label{fig:v12_coord}
\end{figure}

This section verifies that the mock sky coordinates in the light-cone catalog are sufficiently accurate for the pairwise kSZ analysis. The light-cone images are constructed by co-adding images from flat simulation slices\,(see Sect.\,\ref{sec:sim}). The mock galaxy positions in (R.A.,\,Dec.) are derived from pixel coordinates, using the ratio between the image angular size and the number of pixels. For the 5$\times5\,\mathrm{deg}^2$ image, the angular pixel size is 
\[ \theta_{\rm pix}= \frac{\theta=5^\circ}{N_{\rm pix}=4096}\:.\]
Here $N_{\rm pix}$ is the number of pixels in the mock image. Since the pairwise estimation involves computing large separations between galaxies, and we apply the observational pipeline to mock products, it is important to verify the accuracy of this approximation.

We performed a test to compare distances computed using the exact Cartesian positions with those obtained from the approximate mock sky coordinates. The exact distance calculation takes the Cartesian $(X, Y, Z)$ positions of each mock cluster, where $X, Y$ are obtained from the light-cone geometry and $Z$ from the Hubble redshift of each halo. These positions were converted to sky coordinates (R.A.,\,Dec.) in degrees and radial comoving distance $D$ in $h^{-1}$cMpc
\[ D = \sqrt{X^2+Y^2+Z^2} \,,\]
\[ \mathrm{Dec.} = 90 - \mathrm{arccos}\left(\frac{Z}{D}\right)\times\frac{180}{\pi}\,,\]
\[ \mathrm{R.A.} = \mathrm{arctan}\left(\frac{Y}{X}\right)\times\frac{180}{\pi} + 180 \,.\] 
These coordinates are only used for distance calculation, and do not affect the sample selection. The photometry pipeline is not affected by this approximation, since it operates on Plate-Carr\'{e}e projection that is consistent with the mock sky coordinates.

The test was performed on the 23,\,128 mock clusters selected in Sect.\,\ref{subsec:richness}. These clusters were located at redshift $z=0.2-0.6$ and had a minimum mass of $\mathrm{log}\,(M_{200m}/h^{-1}\mathrm{M_{\odot}}) \geq 12.1$. The comparison is shown in Fig.\,\ref{fig:v12_coord}, where the pairwise velocities obtained from the approximated and exact sky coordinates agree well within the uncertainties estimated by bootstrapping the cluster catalog 1000 times. Therefore, the impact of the sky coordinate approximation is negligible.

\FloatBarrier 
\twocolumn

\clearpage

\end{appendix}
\end{document}